\documentclass[apj]{emulateapj}

\slugcomment{To appear in The Astrophysical Journal Supplement Series}

\shorttitle{Optical spectroscopy of X-ray sources}
\shortauthors{Silverman et al.}

\begin{document}



\title{The Extended $Chandra$ Deep Field-South Survey: Optical spectroscopy of faint X-ray sources with the VLT\footnote{Based on observations made with ESO Telescopes at Paranal Observatories under programme IDs 072.A-0139 and 080.A-0411.} and Keck\footnote{``Some of the data presented herein were obtained at the W.M. Keck Observatory, which is operated as a scientific partnership among the California Institute of Technology, the University of California and the National Aeronautics and Space Administration. The Observatory was made possible by the generous financial support of the W.M. Keck Foundation."}}


\author{J. D. Silverman \altaffilmark{1,2,3}, V. Mainieri\altaffilmark{4}, M. Salvato\altaffilmark{5}, G. Hasinger\altaffilmark{5}, J. Bergeron\altaffilmark{6}, P. Capak\altaffilmark{7}, G. Szokoly\altaffilmark{8}, A. Finoguenov\altaffilmark{2}, R. Gilli\altaffilmark{9}, P. Rosati\altaffilmark{4}, P. Tozzi\altaffilmark{10}, C. Vignali\altaffilmark{11}, D. M. Alexander\altaffilmark{12}, W. N. Brandt\altaffilmark{13}, B. D. Lehmer\altaffilmark{14,15}, B. Luo\altaffilmark{13}, D. Rafferty\altaffilmark{13}, Y. Q. Xue\altaffilmark{13}, I. Balestra\altaffilmark{2}, F. E. Bauer\altaffilmark{16,17}, M. Brusa\altaffilmark{2}, A. Comastri\altaffilmark{9}, J. Kartaltepe\altaffilmark{18}, A. M. Koekemoer\altaffilmark{19}, T. Miyaji\altaffilmark{20}, D. P. Schneider\altaffilmark{13}, E. Treister\altaffilmark{18}, L. Wisotski\altaffilmark{21}, M. Schramm\altaffilmark{22}}


\altaffiltext{1}{Institute of Astronomy, ETH Z\"urich, CH-8093, Z\"urich, Switzerland.}
\altaffiltext{2}{Max-Planck-Institut f\"ur extraterrestrische Physik, D-84571 Garching, Germany}
\altaffiltext{3}{Current address: Institute for the Physics and Mathematics of the Universe (IPMU), University of Tokyo, Kashiwanoha 5-1-5, Kashiwa-shi, Chiba 277-8568, Japan}
\altaffiltext{4}{European Southern Observatory, Karl-Schwarzschild-Strasse 2, Garching, D-85748, Germany}
\altaffiltext{5}{Max-Planck-Institute for Plasma Physics, Boltzmannstrass 2, Garching, D-85748, Germany}
\altaffiltext{6}{Institut d'Astrophysique de Paris, 98bis Boulevard, F-75014 Paris, France}
\altaffiltext{7}{California Institute of Technology, 1200 East California Boulevard, Pasadena, CA 91125}
\altaffiltext{8}{E\"otv\"os University, Institute of Physics, 1117 Budapest, P\'azm\'any P. s. 1/A, Hungary}
\altaffiltext{9}{Instituto Nazionale di Astrofisica (INAF) - Osservatorio Astronomico di Bologna, Via Ranzani 1, 40127 Bologna, Italy}
\altaffiltext{10}{INAF - Osservatorio Astronomico di Trieste, via G. B. Tiepolo 11, 34131 Trieste, Italy}
\altaffiltext{11}{Dipartimento di Astronomia, Universit\'a degli Studi di Bologna, Via Ranzani 1, 40127 Bologna, Italy}
\altaffiltext{12}{Department of Physics, University of Durham, South Road, Durham, DH1 3LE, UK} 
\altaffiltext{13}{Department of Astronomy \& Astrophysics, 525 Davey Lab, The Pennsylvania State University, University Park, PA 16802, USA}
\altaffiltext{14}{The Johns Hopkins University, Homewood Campus, Baltimore, MD, 21218, USA}
\altaffiltext{15}{NASA Goddard Space Flight Centre, Code 662, Greenbelt, MD, 20771, USA}
\altaffiltext{16}{Space Science Institute, 4750 Walnut Street, Suite 205, Boulder, Colorado 80301}
\altaffiltext{17}{Pontificia Universidad Cat\'{o}lica de Chile, Departamento de Astronom\'{\i}a y Astrof\'{\i}sica, Casilla 306, Santiago 22, Chile}
\altaffiltext{18}{Institute of Astronomy, University of Hawaii, 2680 Woodlawn Drive, Honolulu, HI, 96822}
\altaffiltext{19}{Space Telescope Science Institute, 3700 San Martin Drive, Baltimore, MD 21218, USA}
\altaffiltext{20}{Instituto de Astronomia, Universidad Nacional Autonoma de Mexico-Ensenada, Km 103 Carretera Tijuana-Ensenada, BC 22860}
\altaffiltext{21}{Astrophysikalisches Institut Potsdam, An der Sternwarte 16, 14482 Potsdam, Germany}
\altaffiltext{22}{Department of Astronomy, Kyoto University, Kyoto 606-8502, Japan}

\begin{abstract}

We present the results of a program to acquire high-quality optical spectra of X-ray sources detected in the Extended-$Chandra$ Deep Field - South (\hbox{E-CDF-S}) and its central 2 Ms area.  New spectroscopic redshifts, up to $z=4$, are measured for 283 counterparts to $Chandra$ sources with deep exposures ($t\sim2-9$ hr per pointing) using multi-slit facilities on both the VLT (VIMOS) and Keck (DEIMOS) thus bringing the total number of spectroscopically-identified X-ray sources to over 500 in this survey field.  Since our new spectroscopic identifications are mainly associated with X-ray sources in the shallower 250 ks coverage, we provide a comprehensive catalog of X-ray sources detected in the \hbox{E-CDF-S} including the optical and near-infrared counterparts, determined by a likelihood routine, and redshifts (both spectroscopic and photometric), that incorporate published spectroscopic catalogs, thus resulting in a final sample with a high fraction (80\%) of X-ray sources having secure identifications.  We demonstrate the remarkable coverage of the luminosity-redshift plane now accessible from our data while emphasizing the detection of AGNs that contribute to the faint end of the luminosity function ($L_{0.5-8~{\rm keV}}\sim 10^{43-44}$ erg s$^{-1}$) at $1.5\lesssim z \lesssim 3$ including those with and without broad emission lines.  Our redshift catalog includes 17 type 2 QSOs at $1 \lesssim z \lesssim 3.5$ that significantly increases such samples ($2\times$).  Based on our deepest (9hr) VLT/VIMOS observation, we identify ``elusive" optically-faint galaxies ($R_{mag}\sim{25}$) at $z\sim2-3$ based upon the detection of interstellar absorption lines (e.g., OII+SiIV, CII], CIV); we highlight one such case, an absorption-line galaxy at $z=3.208$ having no obvious signs of an AGN in its optical spectrum.   In addition, we determine accurate distances to eight galaxy groups with extended X-ray emission detected both by $Chandra$ and XMM-$Newton$.  Finally, we measure the physical extent of known large-scale structures ($z\sim0.7$) evident in the CDF-S.  While a thick sheet (radial size of 67.7 Mpc) at $z\sim0.67$ extends over the full field, the $z\sim0.73$ structure is thin (18.8 Mpc) and filamentary as traced by both AGNs and galaxy groups.  In the appendix, we provide spectroscopic redshifts for 49 counterparts to fainter X-ray sources detected only in the 1 and 2 Ms catalogs, and 48 $VLA$ radio sources not detected in X-rays.

\end{abstract}





\section{Introduction}

It has become clear that the remarkable capability of the $Chandra$
observatory, as demonstrated with the observations of the $Chandra$
Deep Field - North \citep{br01,al03} and South \citep{gi02,luo08}, to detect
faint X-ray sources such as obscured AGN \citep[e.g.,][]{ros02,ba03,sz04,to06,daddi07},
star-forming and normal galaxies \citep[e.g.,][]{al02,horn03,ba04,le07,le08} and extended group/cluster
emission \citep[e.g.,][]{ba02} out to cosmological redshifts must be further
exploited \citep[see][for a review]{br05,br10}.  For instance, the
predominance of moderate-luminosity AGNs ($L_X\sim10^{43}$ erg
s$^{-1}$) at $z\sim1$ \citep[e.g.][]{ha05,ba05} in these surveys may be due to
the emergence of young, rapidly growing supermassive black holes
\citep[e.g.][]{ma04,me08,sha09} or the tail end \citep[e.g.,][]{ba07} of an earlier period of rapid growth.  The spatial clustering of these
low-to-moderate luminosity AGNs \citep{gi05} indicate where SMBHs
reside in the context of the large-scale mass distribution.  To constrain evolutionary models of SMBHs, it is imperative to detect
these moderate-luminosity AGNs in significant numbers over a cosmic
time span similar to that covered by luminous quasars ($1 \lesssim z \lesssim
3$) and on larger spatial scales ($>10$ Mpc) than has yet to be
attained due to the limited area coverage of the deep fields.  The `Extended $Chandra$ Deep Field-South' \citep[\hbox{E-CDF-S};][]{le05}, COSMOS \citep{elvis09,puc09}, and the
'Extended Groth Strip' \citep[EGS;][]{na05,laird09} surveys have been observed by $Chandra$ for this purpose.

The \hbox{E-CDF-S} survey has completed a 1 Ms $Chandra$ Very Large Program that covers a wide area (0.33 deg$^2$; 3$\times$ area of the CDF-S) at depths reaching $\sim1\times10^{-16}$ and $\sim7\times10^{-16}$ ergs cm$^{-2}$ s$^{-1}$ for the 0.5--2.0 and 2--8 keV bands, respectively.  This survey adds four $Chandra$ pointings (ACIS-I; $t_{exposure}$$\sim$250 ksec each), each flanking the deeper central CDF-S \citep{gi02,luo08}, that have generated a catalog of 762 X-ray sources \citep{le05} with 559 falling outside the 2 Ms area \citep{luo08}.  This combination of depth and area is optimized to mitigate cosmic variance \citep{ya05,gi05} and detect moderate-luminosity (i.e., Seyfert-type) AGNs beyond $z$$\sim$1.5 that can only be adequately accounted for by X-ray selection; severe host-galaxy dilution in the optical and strong star formation in the IR hampers standard color selection techniques \citep[e.g.,][]{do07}.  As a bonus, wider area X-ray surveys have the potential to detect rare and interesting sources in the distant universe such as double-peaked broad emission-line quasars \citep{luo09}.

Deeper X-ray data have recently been obtained in the central CDF-S regions by $Chandra$ and XMM-$Newton$.  The 2 Ms $Chandra$ observation of the CDF-S \citep{luo08} has brought the exposure to the equivalent depth of the CDF-N \citep{al03} thus effectively improving upon the limited statistics of faint X-ray sources and spectral characterization of already known brighter sources.  Recently, an additional 2 Ms has been allocated to the field with director's discretionary time to reach an unprecedented flux limit ($f_{0.5-2~ {\rm keV}}\sim 1\times10^{-17}$ erg cm$^{-2}$ s$^{-1}$) for X-ray surveys.  Furthermore, an ongoing program is close to reaching a depth of 3 Ms with XMM-$Newton$ that aims to perform X-ray spectral analysis of a large sample of obscured AGN, including a handful that are Compton-thick.

Both the $Chandra$ and XMM-$Newton$ programs complement the extensive multi-wavelength observations of the CDF-S and its extended area.  The optical imaging includes coverage in the $UBVRI$ from the ESO Deep Public Survey \citep[e.g.,][]{ar01} and the 17 filter COMBO-17 survey \citep{wo04,wo08} that provides accurate photometric redshifts ($\sigma_z \approx 0.03$).  High-resolution imaging is available from the HST Advanced Camera for Surveys (ACS) observations via the GEMS \citep{ri04,caldwell08}, GOODS \citep{gi04} and HST Ultra Deep field \citep[UDF; ][]{beckwith06} projects.  Extensive ground-based, near-infrared imaging campaigns \citep{ol06,gr06,taylor09} have been completed in the J, H and K bands.  Deep $Spitzer$ IRAC \citep{dam09} and MIPS observations have facilitated the study of the relationship between AGNs and their host galaxies (i.e., stellar mass, star formation activity).  Deep radio observations at 20 cm wavelength have been acquired with the Very Large Array \citep{kellermann08,miller08,mainieri08}.  Over 1000 spectroscopic redshifts are available for multi-wavelength selected galaxies via the CDF-S \citep{sz04}, VVDS \citep{lef04}, K20 \citep{mi05}, GOODS \citep[e.g., ][]{po09,balestra10} and MUSYC \citep{tr09a} surveys, although many of the $Chandra$ sources lack spectroscopic redshifts and/or are very faint thus photometric redshifts may have considerable uncertainty.

While ground-based spectroscopic campaigns have contributed significantly to many successful studies of the X-ray source population, they have struggled to acquire spectra of a large fraction of the X-ray sources in the deep fields due to the faint optical magnitudes ($R_{mag}\gtrsim 24$) of the counterparts.  As evident in Figure 6 of \citet{br05}, this has impacted the identification of low-to-moderate luminosity AGN at $z>1.5$, whose optical emission is dominated by their host galaxy, thus presenting a challenge equivalent to the 'redshift desert' encountered by high-redshift galaxy surveys.  Even though photometric techniques have been demonstrated to be successful at these redshifts \citep{zh04,ma05,sa09,luo10}, their errors ($\Delta z$$\sim 0.3$ at $z\sim2$; see Fig. 12 of Salvato et al. 2009) are substantial at faint magnitudes ($24 \lesssim R \lesssim 26$) thus limiting their use where precise distances are required (e.g., spatial correlation analysis). Fortunately, strategies to identify these galaxies spectroscopically have been successfully employed with $8-10$m class telescopes \citep{st04,lilly07,po09}.  These efforts require deep spectroscopic observations with wavelength coverage below 5000~\AA~to identify faint AGN signatures (i.e., emission lines such as Ly$\alpha$, CIV and CIII]) or absorption features usually attributed to the interstellar medium in the host galaxies.  The interstellar absorption lines of SiII, OI, OII+SiIV, CII, SiIV, CIV, FeII and AlII will be available within the observable window (3800-6700\AA) over $z \sim 1.5-3.0$ \citep[See Figure 2 of ][]{sh03}.  The detection of Ly$\alpha$ in emission or absorption is an important diagnostic at $z>2$.

In this paper, we present results from our optical spectroscopic program of $Chandra$ sources, mainly in the \hbox{E-CDF-S}, but also including those lacking redshifts in the central CDF-S, with the Very Large Telescope (VLT) and Keck.  We have measured redshifts for 248 X-ray sources from the catalog of \citet{le05} that had been previously unidentified. In total, there are now 422 out of 762 X-ray sources that have a spectroscopic redshift.  We organize the paper as follows.  In Section~\ref{ids}, we provide specific details about the determination of reliable optical and near-infrared counterparts to X-ray sources in the \hbox{E-CDF-S}.  Section~\ref{spectroscopy} describes the processing and analysis of our VLT and Keck spectroscopic observations including our redshift measurements and classification scheme.  In Section~\ref{compilation}, we present a comprehensive catalog of E-CDF-S X-ray sources that have spectroscopic redshift measurements not only from our own data but those previously reported by various programs.  We include newly-derived photometric redshifts (Section~\ref{photoz}), by our team \citep{ra10} that incorporate near-infrared photometry from $Spitzer$ and deeper ground-based imaging \citep[e.g.,][]{taylor09}, for X-ray sources that lack a secure spectroscopic redshift.  In Section~\ref{ecdfs}, we discuss the X-ray and optical properties of the point sources and briefly describe the populations detected in these $Chandra$ observations.  In Section~\ref{lss}, we use our compilation of AGNs to measure the physical extent of the large-scale structures ($z\sim0.7$) found in the CDF-S \citep{gi03} with better precision, and detect additional structures at higher redshifts.  To explore the relationship between AGNs and galaxy groups within these large-scale structures,  we report on the identification of extended sources, primarily X-ray emission associated with galaxy groups and clusters based on both the $Chandra$ and XMM-$Newton$ observations of the \hbox{E-CDF-S}.  Furthermore, we present in the appendix additional spectroscopic redshifts for targets from our spectroscopic program including counterparts to 49 X-ray sources solely detected in the 1--2 Ms catalogs \citep{gi02,luo08} and 48 $VLA$ radio sources not detected with the current X-ray data.  

Throughout this work, we assume H$_{\circ}=70$ km s$^{-1}$ Mpc$^{-1}$,
$\Omega_{\Lambda}=0.7$, and $\Omega_{\rm{M}}=0.3$.  All magnitudes are expressed in the AB system where $R_{\rm AB}=R_{\rm Johnson} +0.055$.

\section{Optical and near-infrared identification} 
\label{ids}

We utilize published optical and near-infrared catalogs, listed in Table~\ref{catalogs}, that cover the \hbox{E-CDF-S} to identify counterparts to the 762 $Chandra$ sources presented in \citet{le05}.  A higher priority is given to catalogs based on imaging with superior resolution and depth at red optical wavelengths thus our weighting is set accordingly (HST/ACS $i$-band, WFI $R$-band, ISAAC K$_{\rm s}$ SOFI K$_{\rm s}$).  A likelihood-ratio technique \citep[e.g.,][]{su92,ru00,ci03,ci05,br07} is implemented, since there can be multiple possible counterparts within an X-ray error circle mainly due to the faint ($R\sim26$) optical depths required to identify the majority of the X-ray sources.  This method provides a quantitative measure of reliability for each counterpart essentially based on its angular distance from the X-ray centroid ($r$) and its magnitude ($m$).  Specifically, the likelihood ratio (LR; equation~\ref{eq:lr}) is the probability that a given optical or near-infrared source is the true counterpart to the X-ray detection, relative to the chance that the same object is a random background source.

\begin{deluxetable}{lll}
\tabletypesize{\scriptsize}
\tablecaption{Public imaging catalogs\label{catalogs}}
\tablewidth{0pt}
\tablehead{
\colhead{Name}&\colhead{Band}&\colhead{Reference}
}
\startdata
WFI&R&\citet{ar01}\\
GEMS&v,z &\citet{caldwell08}\\
GOODS/ISAAC&J, H, Ks&Retzlaff et al., in prep\\
EIS/SOFI&J, H, Ks&\citet{ol06}\\
\enddata
\end{deluxetable}

\begin{equation}
\label{eq:lr}
LR = \frac{q(m)f(r)}{n(m)}
\end{equation}

\noindent The function $q(m)$ describes the probability of a source of magnitude $m$ being the likely counterpart;  this probability distribution is generated by comparing the magnitude distribution of optical counterparts within a specified radius of the X-ray sources to the distribution of background objects.  Since we expect all true counterparts to be within a radius of $2\arcsec$ from their corresponding X-ray centroids, the distribution function {\it q(m)} is determined by selecting only sources within this radius.  The probability distribution based on the radial offset $f(r)$ is modeled as a Gaussian function with a variance that depends both on the X-ray and optical uncertainties taken in quadrature.  We assume a fixed radial uncertainty in the position of the optical/near-infrared counterparts to be $0.5\arcsec$ ($\sim3\sigma$).  The likelihood ratio is normalized by the surface density {\it n(m)} of objects clearly not associated with the X-ray counterparts (i.e., background objects).  Further explicit details on the determination of these three distribution functions can be found in \citet{ci03}.  We further require the likelihood ratio LR for each candidate to be above 0.2 and visually inspect each case to generate a final catalog of high confidence optical/near-infrared counterparts.  The latter step enables accurate identification of counterparts in a few specific cases where there may be confusion between nearby faint and bright optical candidates, crowded regions, or sources with complex optical morphologies.
  
We then measure the reliability \citep[$0<{R_j}<1$; ][]{su92}
that a particular source ($j$) is the true counterpart.

\begin{equation}
R_j = \frac{LR_j}{\sum_i LR_i + (1-Q)}
\label{eq:reliable}
\end{equation}

\noindent The sum is over all counterparts ($i$) associated with
a single X-ray source.  The probability that a counterpart is above
the magnitude limit of our optical/near-IR catalog is taken into
consideration (Q$=\int_{- \infty}^{m_{lim}}q(m)~dm)$.  Here, we fix
Q$=0.8$ that corresponds to the ratio of the expected number of
identifications (the integral of {\it q(m)}) to the total number of
X-ray sources.  We note that our results are insensitive to
differences in $Q$ within the range of 0.5-1.0.

\begin{figure}
\epsscale{1.2}
\plotone{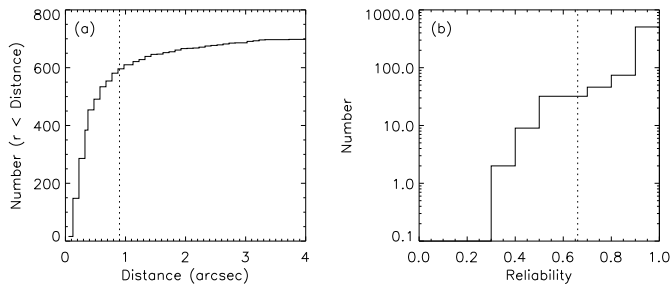}
\caption{($a$) Cumulative number of X-ray sources as a function of
the distance to their most likely optical/near-infrared
counterpart. The vertical line marks the distance which includes 90\% of
the sources .  ($b$) The distribution of the reliability
index ($R$) for the likely counterparts.  The value of $R$ for which 90\% of the optical/near-infrared counterparts have a higher reliability is shown by the dashed line.}
\label{xo_match}
\end{figure}

Our final catalog contains counterparts to 677 X-ray sources.  We find that 90\% of the counterparts have offsets less than $0.9\arcsec$ (Figure~\ref{xo_match}a) and have a reliability parameter $R>0.68$ (Fig~\ref{xo_match}b).  Only 36\% of the X-ray sources have a near-infrared counterpart in the K$_{\rm s}$ band \citep[$K_{sAB}^{lim}=21.4$;][]{ol06}, thus highlighting the importance of deeper observations \citep[$K_{sAB}^{lim}\sim22$; ][]{taylor09} in this filter.  For example, \citet{brusa2010} find counterparts to nearly 100\% of the X-ray sources in the COSMOS field  due to a deep $K_{sAB}$ band photometry \citep{mccracken2010} coupled with the fact that the XMM-$Newton$ sources in the COSMOS field are brighter in comparison to X-ray sources in the $Chandra$ Deep fields.  We stress that our main objective in the present study is to provide optical spectroscopy of faint optical counterparts and thus recent improvements with respect to the depth of public near-infrared imaging is not the focus of the present work.  Highly complete catalogs of counterparts to X-ray sources based on deeper near-infrared imaging can be found elsewhere such as \citet{cardamone08} who find a  higher detection rate ($\sim90\%$) when incorporating the $Spitzer$/IRAC observations of the \hbox{E-CDF-S}.  Our final catalog is presented in Section~\ref{compilation}.

\section{Optical spectroscopy}
\label{spectroscopy}

There are over one thousand spectroscopic redshifts available in the CDF-S region.  The majority have been supplied by galaxy redshift surveys on the VLT such as VVDS \citep{lef05}, GOODS \citep{va05,va06,va08,po09,balestra10}, and K20 \citep{mi05}.  Most relevant for our study is the redshift survey of X-ray sources in the 1 Ms CDF-S \citep{sz04} that provides redshifts for 74 sources in the \citet{le05} catalog.  At the outset of our program, there were published spectroscopic redshifts for only $\sim 100$ \hbox{E-CDF-S} sources, thus justifying the need for a dedicated spectroscopic program specifically targeting the X-ray source population. We note that an independent effort similar in scope to our program has been undertaken \citep{tr09a} on Magellan and the VLT that acquired spectra of shallower depth and at red wavelengths; these spectroscopic redshifts are incorporated into our catalog (Section~\ref{compilation}) if there is confidence that an improvement in the quality of the overall catalog is assured.  In addition, we include newly available redshifts provided by the GOODS program \citep{balestra10}.  As further described below, we have obtained optical spectroscopic observations with both the VLT and Keck of the faint X-ray source population detected not only in the \hbox{E-CDF-S} but also in the 1-2 Ms catalogs \citep{gi02,luo08}.  Additional slits have been placed on galaxies associated with X-ray selected clusters and non-X-ray targets such as VLA radio sources, $Spitzer$ IRAC/MIPS sources, and galaxies ($R_{\rm AB}<25$) in the GOODS/MUSIC catalog \citep{gr06} having photometric redshifts between $1.5<z<2.5$.  All processed flux-calibrated spectra, in FITS and ASCII formats, are available
online\footnote{\url{http://member.ipmu.jp/john.silverman/CDFS.html and http://www.eso.org/\textasciitilde vmainier/cdfs\_pub/}} with the exception of the non-active galaxies (i.e., GOODS/MUSIC, $Spitzer$ MIPS and IRAC) that will be presented separately with publicly available spectra.

\begin{figure}

\includegraphics[width=9cm]{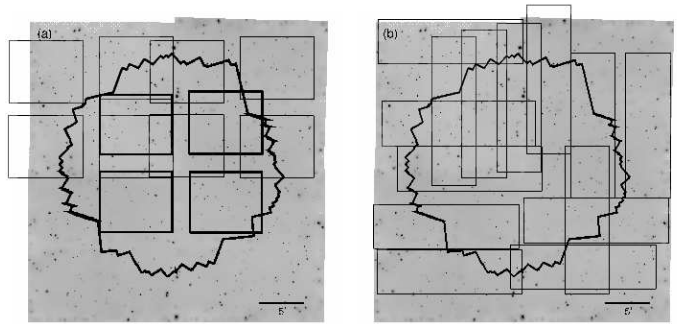}

\caption{Footprints of our multi-aperture spectroscopic observations
overlaid onto a smoothed X-ray image of the \hbox{E-CDF-S}.  ($a$) The four
quadrants of our three VLT/VIMOS pointings are marked in black with
the central deep (9hr) exposure highlighted by thick lines.  ($b$)
Fourteen Keck/DEIMOS pointings are shown by the black rectangular boxes.
North is up and East is to the left.  In both panels, the jagged region
marks the central 2Ms $Chandra$ region.}

\label{spec_prog}
\end{figure}

\subsection{VLT/VIMOS observations, data reduction \& analysis}

We have acquired optical spectra with the VLT, using the Visible Multi-Object Spectrograph \citep[VIMOS;][]{lef03} mounted on the Nasmyth focus of UT3 Melipal.  Our campaign, in service mode, primarily targeted counterparts to $Chandra$ X-ray sources in the \hbox{E-CDF-S} that lacked spectroscopy.  VIMOS is ideal for spectroscopy of optically faint targets due to its multi-slit capability over a wide field of view that is well matched to $Chandra$ since an area of $7\arcmin\times8\arcmin$ is covered by each of 4 CCDs.

We have carried out two pointings, each with two different setups, in the northern half of the \hbox{E-CDF-S} field (Figure~\ref{spec_prog}$a$).  In many cases, we obtain spectra for each X-ray source covering a wide wavelength range to detect spectroscopic features in a variety of AGNs;  this requires the use of two different dispersive elements separately.  The difference in spectral resolution (as described below) essentially prevents us from observing all sources with both configurations.  Since many of these AGN are obscured and reside in early-type galaxies at $0.4<z<1.2$, we used the medium-resolution (MR) grism in the red that allows a better subtraction of night-sky lines and has higher transmission compared to the low-resolution (LR) red grism.  The MR grism has the following characteristics: a wavelength range of $5500-9600~\rm{\AA}$, $R\sim580$ ($\Delta v \sim 500$ km s$^{-1}$) and dispersion of 2.5 ${\rm \AA}$ pix$^{-1}$.   To cover shorter wavelengths (3700-6700 ${\rm \AA}$), we carry out observations using the low-resolution blue grism (LR-blue; $R=180$, dispersion=5.3 ${\rm \AA}$ pix$^{-1}$ ) to detect emission lines (i.e., Ly$\alpha$, CIV, CIII]) in AGN with $1.5<z<4$ or absorption lines mainly attributed to the ISM of their host galaxy when no AGN signatures are present.  Observations with the LR-blue grism are crucial to identify the obscured AGN at these redshifts that are underrepresented in the identification of sources in Deep $Chandra$ surveys \citep[e.g., See Figure 6 of][]{br05}.  A wide spectral range greatly facilitates object classification, since many of these optically underluminous AGN suffer from severe host dilution.  It is worth highlighting that expanded wavelength coverage can detect broad optical emission lines in sources previously classified as normal or narrow emission-line galaxies \citep[e.g., ][]{pa03,si05}.  Our total exposure times ($\sim5$ hours per mask) are set to identify faint optical counterparts to a limiting magnitude of $R\approx25$ for AGN with strong emission lines and $R=24$ for highly obscured AGN having emission/absorption features of moderate strength.

A single, deep VIMOS exposure was obtained in the central CDF-S to identify the obscured AGN population at $1.5<z<3$.  The setup was configured with the LRblue grism and OS-blue filter that provided a wavelength coverage sufficiently blue to identify spectral features in galaxies at these redshifts.  Initially, two separate pointings were scheduled with integration times of 18 hours each, but scheduling difficulties resulted in a single pointing with a total on-source exposure time of 9hr.  The observation was split into 18 segments each having an exposure time of 30 minutes.  This single observation is unique due to its depth and enables us to achieve goals such as identifying the obscured AGN at $z>1.5$ that tend to have faint optical counterparts ($R\sim25$) and lack spectroscopic identification.

All masks were designed with the VIMOS Mask Preparation Software \citep[VMMPS; ][]{bo05} that optimizes the slit assignments based on our input catalog and VIMOS pre-imaging.  The coordinates of our targets, initially selected from our deep $R$-band WFI image, were transformed to the reference frame of the pre-image.  Optical counterparts to X-ray sources are designated as ``Compulsory'' in the input catalog and are given top priority;  additional targets, as mentioned above, are labelled as ``Selectable''.  All slits had a width of 1\arcsec.  Two stellar-like objects were milled for each of the four quadrants for field acquisition.

We list the details of the VLT observations in Table~\ref{vltlog}.  Science exposures of 955-1800 seconds each were dithered by $\sim1.5\arcsec$ while keeping the object in the slit for optimal removal of night-sky emission.  Actual exposure times per mask range from 3.2 to 9 hours.  The typical seeing was $\sim$1$\arcsec$  and airmass $\lesssim1.2$.  All necessary calibration frames (e.g., dome flats, arc lamps) are taken including short spectroscopic observations of standard stars required for adequate flux calibration. The data were reduced using the VIMOS Interactive Pipeline and Graphical Interface \citep[VIPGI;][]{sc05} software.

\begin{deluxetable}{llllc}
\tabletypesize{\scriptsize}
\tablecaption{Log of VLT/VIMOS observations\label{vltlog}}
\tablewidth{0pt}
\tablehead{
\colhead{Date}&\colhead{RA}&\colhead{DEC}&\colhead{Grism\tablenotemark{a}}&
\colhead{Exposure time\tablenotemark{c}}\\
\colhead{UT}&\colhead{J2000}&\colhead{J2000}&&sec (\# of exp.)}
\startdata
2004-10-21&03:31:59&-27:42:23&LRblue&5730.0 (6)\\
2004-10-23&''&''&LRblue&955.0 (1)\\
2004-11-11&''&''&LRblue&8595.0 (9)\\
2004-11-13&''&''&MRred&2865.0 (3)\\
2004-12-03&"&"&MRred&2865.0 (3)\\
2004-12-04&''&''&MRred&2865.0 (3)\\
2004-12-05&''&''&MRred&2865.0 (3)\\
2004-12-07&''&''&MRred&5730.0 (6)\\
2004-12-08&''&''&MRred&2865.0 (3)\\
2005-01-02&03:33:10&-27:42:28&MRred&2865.0 (3)\\
2005-01-03&''&''&MRred&1910.0 (2)\\
2005-01-05&''&''&MRred&1910.0 (2)\\
2005-01-06&''&''&MRred&9550.0 (10)\\
2005-10-04&''&''&LRblue&5730.0 (9\tablenotemark{c})\\
2005-10-30&''&''&LRblue&5730.0 (6)\\
2007-12-12&03:32:25&-27:48:36&LRblue& 7200 (4)\\
2007-12-13&"&"&LRblue&3600 (2)\\
2007-12-14&"&"&LRblue&3600 (2)\\
2007-12-15&"&"&LRblue&3600 (2)\\
2008-01-04&"&"&LRblue&3600 (2)\\
2008-01-06&"&"&LRblue& 3600 (2)\\
2008-01-07&"&"&LRblue& 3600 (2)\\
2008-02-01&"&"&LRblue& 3600 (2)\\
\enddata
\tablenotetext{a}{The filter OS-blue was used in conjunction with the LR
observations and the GG475 for the MR.}
\tablenotetext{b}{The exposure time is the sum of all science frames.}
\tablenotetext{c}{Three observations were not used in the reduction due to a large pixel shift that could not easily be
rectified within the VIPGI environment.}
\end{deluxetable}

\subsection{Keck/Deimos spectroscopy}

Optical spectra were obtained in January of 2007, 2008 and 2010 on Keck II with the Deep Imaging Multi-object Spectrograph \citep[DEIMOS;][]{fa03}.  These observations have a substantially higher spectral resolution than those with the VLT.   In Figure~\ref{spec_prog}$b$, we show the placement of fourteen slit masks with respect to the \hbox{E-CDF-S} field.  For the most part, we use the 600 l mm$^{-1}$ grating and a blue blocking filter (GG455) to achieve a wavelength coverage of 4600-9700 {\rm \AA}, a dispersion of 0.65 {\rm \AA} pix$^{-1}$ and spectral resolution of $\Delta\lambda_{FWHM}\approx3.5$ {\rm \AA} ($\approx 125$ km s$^{-1}$).  Although, six masks were observed with a slightly redder filter (GG495), and a single observation was performed using the 830 l mm$^{-1}$ grating due to a limitation on having multiple configurations in a single evening since the primary target during most of the later half of the night was the COSMOS field.  This higher resolution observation has a wavelength coverage of 5700-8000 ${\rm \AA}$, a dispersion of 0.47 {\rm \AA} pix$^{-1}$ and spectral resolution of $\Delta\lambda_{FWHM}\approx2.5~{\rm \AA}$.  All slits were milled with a width of 1\arcsec.  Integration times ranged from 1 to 3.6 hours per mask.  Wavelength calibration is achieved using short arc lamp exposures.  Observations of a single standard star in the same evening and dome flats allow flux calibration of our spectra.  Due to the southern location of the CDF-S, all observations are carried out at high airmass (1.5-2.0), although under photometric conditions.  In Table~\ref{kecklog}, we provide a log of the observations including some relevant details.

All data reduction was performed using the DEEP2 software pipeline.\footnote{Further details can be found at \url{http://www2.keck.hawaii.edu/inst/deimos/pipeline.html}}  This is essentially an automated routine that identifies slit positions, determines the two-dimensional wavelength solution based on the well-understood geometrical distortions of the instrument, removes cosmic rays, and traces the targets to provide one-dimensional, sky-subtracted spectra.  In a few specific cases mainly pertaining to standard-star observations, we manually extracted the spectra with a modified version of the DEEP2 script ('do\_extract').  Flux calibration is then applied to all 1-D FITS spectra using
IRAF.\footnote{IRAF is distributed by the National Optical Astronomy Observatories, which are operated by the Association of Universities for Research in Astronomy, Inc., under cooperative agreement with the National Science Foundation.}

\begin{deluxetable}{llllll}
\tabletypesize{\scriptsize}
\tablecaption{Log of Keck/DEIMOS observations\label{kecklog}}
\tablewidth{0pt}
\tablehead{
\colhead{Date}&\colhead{RA}&\colhead{DEC}&\colhead{Grism}&\colhead{Filter}
&\colhead{Exp time}\\
\colhead{UT}&\colhead{J2000}&\colhead{J2000}&l/mm&(GG)&sec(\# exp)}
\startdata
2007-01-21&03:32:59.9&-27:45:10&600 &455&3600 (3)\\
2007-01-18\tablenotemark{a}&03:31:55.0&-27:44:25&600 &495&3600 (3)\\
2007-01-14&03:33:06.8&-27:56:49&600&455& 7200  (4)\\
2007-01-17&03:33:03.2&-28:01:48&600&455& 9000  (5)\\
2007-01-15&03:31:50.1&-27:55:58&600&455& 9000  (5)\\
2007-01-16&03:31:57.5&-28:01:10&600 &455&9000  (5)\\
2008-01-08&03:32:54.0&-27:50:00&830 &495&6000  (5)\\
2008-01-09&03:33:04:0&-27:35:19&600 &495&4800 (5)\\
2008-01-10&03:31:28.2&-27:44:05&600 &495&3600 (3) \\
2010-01-14&03:32:28.9&-27:41:15&600 &495&9000 (5)\\
2010-01-15&03:32:13.8&-27:39:09&600 &495&9000 (5)\\
2010-01-16&03:32:50.5&-27:41:58&600 &495&9000 (5)\\
2010-01-17&03:33:02.0&-27:42:43&600 &495&8707 (5)\\
2010-02-11&03:31:54.2&-27:54:59&600 &495&13016 (11)\\
\enddata
\tablenotetext{a}{Masks with position angle equal to 0 (N-S). All others are oriented in the E-W direction.}
\end{deluxetable}

\subsection{Redshift measurements and source classification}
\label{class}

\begin{figure*}
\hspace{1cm}
\includegraphics[angle=0,width=16cm]{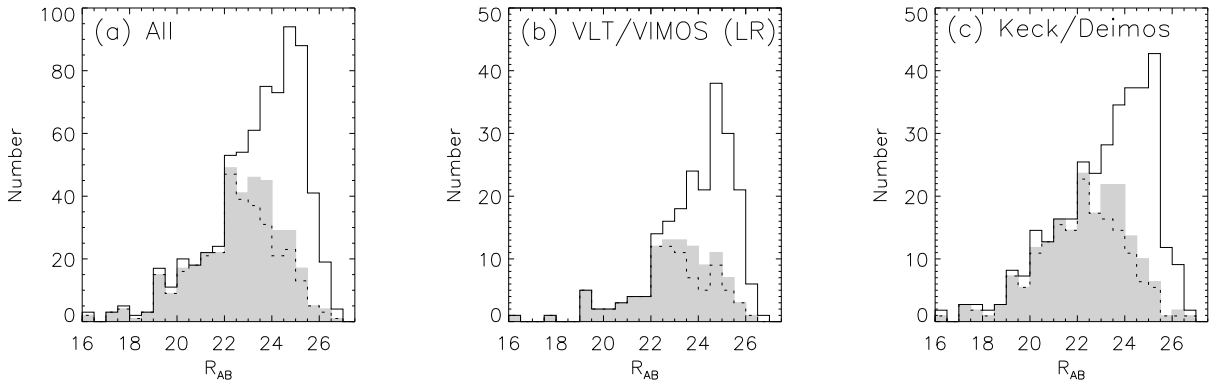}
\caption{Number distribution of all observed X-ray counterparts (open histogram) and those with reliable spectroscopic redshifts ($Flag \ge1$: filled histogram; $Flag=2$: dashed line):
(a) all, (b) VLT-VIMOS (LR blue only), (c) Keck/DEIMOS.}
\label{spectro_eff}
\end{figure*}

We measure redshifts using the KBRED and EZ\footnote{\url{http://cosmos.iasf-milano.inaf.it/pandora/EZ.html}} software that cross-correlates each spectrum with a range of spectral templates.  KBRED and EZ have been developed by the VVDS consortium to optimize the analysis of large numbers of galaxy spectra being generated from large surveys (e.g., VVDS, zCOSMOS).  These templates include stars, quasars, and galaxies with a varying degree of emission/absorption line strength over the wavelength range 1100--9000 $\rm{\AA}$.  High-redshift ($z>1.5$) templates have been provided by the zCOSMOS and VVDS teams that were generated from a sample of galaxies using a similar observing configuration (i.e., spectral resolution, wavelength coverage) with the LRblue grism (see \citealt{lilly07} for further details). Specifically, these templates aid in the identification of galaxies with redshifts between $1.5<z<3.0$ that are known to be observationally challenging.  Each redshift is visually inspected in the VIPGI environment; in some cases the input parameters such as wavelength range or template must be further restricted.  For spectra with strong, narrow emission lines, we use VIPGI to determine the centroid of each line and then compute a mean redshift.  We assign a quality flag to each redshift measurement as defined here:

\begin{itemize}
\item Flag 2: High-confidence redshift ($>95\%$ secure) due to high S/N spectra and multiple spectral features 
\item Flag 1: Redshift with a significant amount of uncertainty at the $\sim50\%$ level due to either low S/N spectra or the presence of only a single emission line.         
\item Flag 0: No redshift measurement is achieved.
\end{itemize}

It is worth highlighting that the assignment of a redshift with a quality flag of 1 is a judgement call that reflects a situation where a solution is likely to be correct but has considerable uncertainty due to usually either poor signal-to-noise, a lack of spectral features (i.e., emission and absorption lines) or both.  A single emission line present in a spectrum with a faint continuum can be attributed to Ly$\alpha$, CIII], MgII, [OII] or H$\alpha$.  We note that it is not possible to empirically determine our success rate with respect to redshift measurements since a coordinated effort to carry out repeat observations has not been undertaken.    Although, as described below (Section~\ref{photoz}), we do compare our spectroscopic redshifts with low quality flags to their photometric estimates and find that $\sim50\%$ are likely to be correct.    

We have measured new redshifts (flag $\ge$ 1) for counterparts to 283 X-ray sources detected in the combined \citet{le05}  and \citet{luo08} catalogs.  To demonstrate our success rate, we show in Figure~\ref{spectro_eff} the optical magnitude distribution of all X-ray sources targeted by our VLT and Keck observations and highlight those for which we have obtained a redshift (Filled histogram: $flag\ge1$; dashed line: $flag=2$).  These distributions represent every attempt to measure a redshift based on a slit being placed on a counterpart to a X-ray source.  Therefore, a single target may be counted more than once depending on how many attempts were made to acquire an optical spectrum that provides a reliable redshift.  For instance, a spectrum is likely to be acquired using Keck if the VIMOS observation did not yield a secure redshift.  As clearly evident (Fig.~\ref{spectro_eff}$a$), we have a high success rate of 83.2\% ($flag \ge 1$) for bright counterparts ($R_{AB}<24$) that drops to 75.1\% for the most secure cases ($flag=2$).  At $R_{AB} \sim24$, our success rate drops to $\sim$50\%; this is roughly a magnitude fainter than the success rate achieved with previous campaigns \citep{sz04,tr09a} in the \hbox{E-CDF-S}.    It is worth highlighting that the sharp cutoff in the success rate with Keck/DEIMOS for $R_{AB}>24$ (Fig.~\ref{spectro_eff}$c$) is significantly less severe for our deeper ($t_{exp}\sim5$ or 9 hours) observations using the LR grism on VLT/VIMOS (Fig.~\ref{spectro_eff}$b$) thus resulting in a number of identifications at these faint magnitudes that attests to the effectiveness of our strategy;  even at the extreme limit for spectroscopy ($R_{AB}\sim26$), we also have a handful of identifications.

The number of independent spectroscopic programs in the CDF-S enables us to further improve the quality of the catalog.  For example, there are 132 optical counterparts to X-ray sources in the \hbox{E-CDF-S} that have multiple redshift measurements. Of these, there are 35 cases that are clearly discrepant.  The majority of these can be explained by the varying quality of the spectra.  For example, the cases in common (24) with the MUSYC program \citep{tr09a} are mainly due to a significant uncertainty based on their low-quality flags for these objects most likely due to lower signal-to-noise spectra acquired from smaller aperture telescopes (i.e., Magellan) than employed in this study.  We have reconciled with the MUSYC team a handful of cases (8), for which both teams measured a different redshift and assigned a 'secure' quality flag, by inspecting all available spectra;  the conclusion was that these eight MUSYC redshifts were incorrect.  Also, three cases of discrepant redshifts with the recent GOODS/VIMOS program \citep{balestra10} were resolved by our higher quality flags and agreement with photometric redshifts.  

Furthermore, we classify each object by the presence of spectral features that fall within the observed optical window as listed here:

\begin{itemize}

\item Broad emission-line AGN (BLAGN): Presence of at least one emission line having $FWHM > 2000$ km s$^{-1}$ 

\item Narrow emission-line galaxy (NELG): Presence of at least one emission line having $FWHM < 2000$ km s$^{-1}$ and no BLAGN signatures.   

\item Absorption-line galaxy (ALG): No emission lines with an observed EQW $>$ 5 ${\rm \AA}$.  Clear signs of absorption lines such as Ca H+K, MgII or G band.

\item Star

\end{itemize}

We note that there are a handful of objects where the classification depended on the observed spectral window.  For example, we initially classified the optical spectra of a few cases as ALG or NELG based on VIMOS/MR coverage with observed wavelength above 5500 ${\rm \AA}$, but we changed the classification to BLAGN due to the presence of broad MgII falling at shorter wavelengths in VIMOS/LR spectra.

\begin{figure*}
\hspace{2cm}
\includegraphics[angle=0,width=14cm]{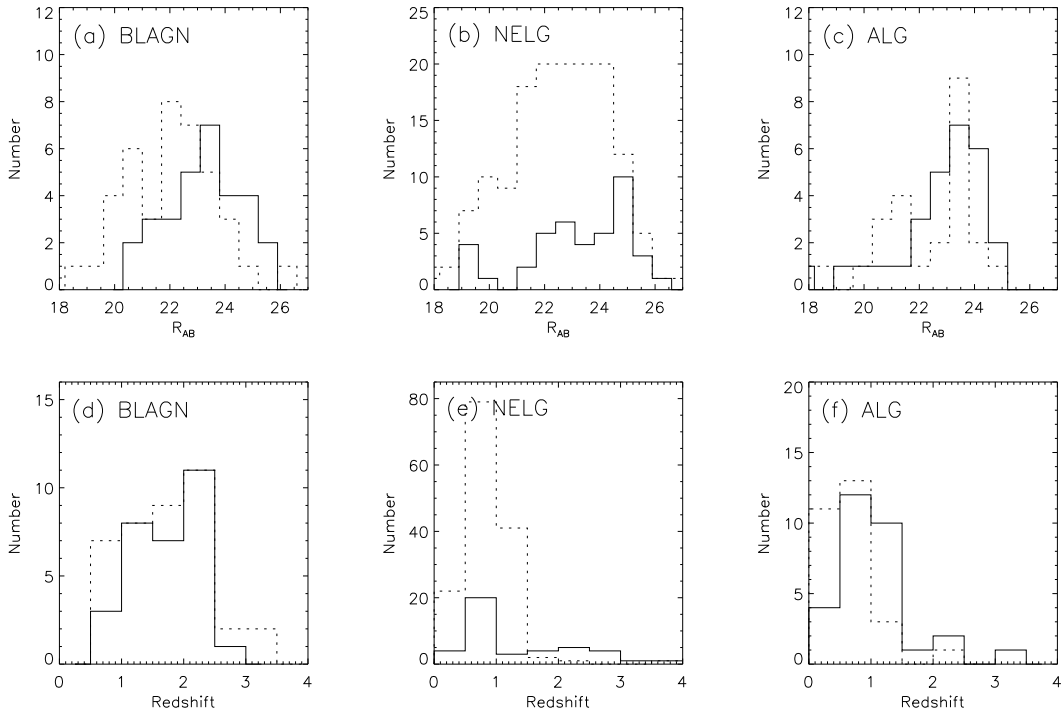}

\caption{Magnitude and redshift distribution of successes ($flag \ge 1$) as shown by their optical classification (BLAGN, NELG, ALG).  The histograms represent the distribution of each quantity and object type for different instrument setups (solid=VLT/VIMOS-LRblue; dashed: Keck/DEIMOS).  We highlight that deep exposures (solid histogram) that cover the blue part of the spectrum result in a shift of the distribution to higher redshift for non-BLAGN.}      
\label{complete_type}
\end{figure*}

Here, we further demonstrate that the observed spectral window not only impacts source classification but the overall redshift distribution of specific types of objects.  In Figure~\ref{complete_type},  we show the optical-magnitude and redshift distributions of the various classes of objects (BLAGN, NELG, ALG) that have fairly secure redshifts ($flag\ge 1$) from our program.  We split the sample by those with redshifts from either the VLT/VIMOS using the LR grism (solid histogram) that is more sensitive at blue wavelengths or Keck/DEIMOS (dashed histogram).  We stress that these distributions are meant to illustrate the success rate of redshift determination and source classification for X-ray sources (from both the E-CDF-S and the central 2 Ms survey) using these two observational configurations.  Therefore, as in Figure~\ref{spectro_eff}, an object can be included in two categories (e.g., ALG and NELG) if different spectra cover different spectral regions.  For example, the X-ray source ECDFS \#411 is included here as an ALG based on a VLT/VIMOS spectrum and also as an NELG from Keck observations with the final classification determined to be NELG.  We see that our deep exposures with VLT/VIMOS (5 or 9 hours) have provided the spectra to identify X-ray sources down to faint optical magnitudes ($R_{mag}\sim25.5$) for all classes (Fig.~\ref{complete_type}$a-c$) as designed.  In the bottom row, we show that the redshift distribution, while similar for BLAGN (Fig.~\ref{complete_type}$d$) between the VLT and Keck observations, is dissimilar for non-BLAGN (Fig.~\ref{complete_type}$e-f$).  Both NELG and  ALG exhibit a more pronounced high redshift component ($1.5\lesssim z \lesssim3$) due to the bluer wavelength coverage that effectively enables the detection of emission lines such as Ly$\alpha$ and CIV, or interstellar absorption lines thus beginning to fill in the 'redshift desert'.  We do recognize that the target selection between the two strategies differs slightly and may bias this comparison in the sense that the VIMOS observations were tuned towards fainter objects.  For example, the VIMOS mask preparation software does not allow one to tilt slits as was done for Keck.  Therefore, in a number of cases, a fainter source was preferred over a brighter target where slit assignment conflicts occurred.  This was especially the case for the single deep (9hr exposure) VIMOS pointing.  Another contributing factor is that many of the brighter targets were assigned redshifts based on prior Keck observations and then subsequently removed from the input catalog of the deep VIMOS pointing.  Such biases are likely responsible for slightly different magnitude distributions shown in Figure~\ref{spectro_eff}.  In consideration of these selection effects, we have compared the redshift distributions as shown in Figure~\ref{complete_type}$e-f$ for both NELGs and ALGs using a restrictive magnitude range ($24<R_{mag}<26$) and still find that the VIMOS LR blue program results in a redshift distribution that is shifted to higher values.                                      

\section{Optical/near-infrared catalog of X-ray sources in the E-CDF-S}
\label{compilation}

In Table~\ref{opt_catalog}, we present a catalog of the most-likely optical/near-infrared counterparts for each X-ray source listed in \citet{le05} and provide parameters (i.e., likelihood ratio, reliability, and angular separation) that quantify the robustness of these multi-wavelength associations.  The spectroscopic information provided here is based on a compilation of all such observations undertaken in the \hbox{E-CDF-S}.  In addition to the spectroscopic redshift, we have included a quality flag that gives the confidence in the redshift measurement.  We attempt to place all quality flags given in the literature on a common scale.  The flags given in \citet{sz04} are translated as follows: a `2' or a `3' is converted to a `2' in our scale, a `1' is kept as a `1', and all others are replaced with a `0'.  With respect to the GOODS-S catalogs \citep{va08,po09,balestra10}, we convert an `A' or `B' into a `2', a `C' becomes a `1' and all others have a `0'.  Quality flags associated with the MUSYC redshifts \citep{tr09a} have been incorporated as provided through private communication on the same scale (0,1,2) as used in this work.  \noindent In addition, a spectral classification is given mainly based on the presence of emission lines and their width.  Finally, a photometric redshift measurement is provided as determined by \citet{ra10} as further described in Section~\ref{photoz}.           

The columns of Table~\ref{opt_catalog} are as follows.

\begin{itemize}

\item{Col. 1~~~~ XID from \citet{le05}}
\item{Col. 2-3~ Right Ascension and Declination (J2000) of the optical/near-infrared counterpart.}
\item{Col. 4~~~~ Catalog associated with the identified counterpart: {\it
      GEMS-$z$} ($z$-band HST/ACS catalog), {\it WFI-R} (WFI R-band
    catalog), {\it SOFI-K} (K-band catalog from SOFI)}, {\it SOFI-H} (H-band catalog from SOFI)
\item{Col. 5~~~~ Separation, in arcsec, between the X-ray source
    and its counterpart.}
\item{Col. 6~~~~    Likelihood ratio (LR) as defined in Section \ref{ids}.}
\item{Col. 7~~~~    Reliability parameter (R) as defined in Section \ref{ids}.}
\item{Col. 8~~~   R-band magnitude (AB).}
\item{Col. 9~~~  Spectroscopic redshift of the counterpart.}
\item{Col. 10~~~ Quality flag for the spectroscopic redshifts: `2' secure, `1' some uncertainty, `0' no redshift; +0.5 if in agreement with photometric redshift}
\item{Col.11}~~~Spectroscopic catalog (1=\citealt{sz04}; 2=FORS2/GOODS; 3=VVDS; 4=New VLT;  5=VIMOS/GOODS; 6=K20; 7=New Keck; 8=\citealt{ravi07}; 9=New VLT/9hr; 10=\citealt{tr09a}; 11=Wisotzki/Schramm, private communication; 15).
\item{Col. 12~~~ Spectroscopic classification: BLAGN, NELG, ALG, STAR}.
\item{Col. 13~~~ Photometric redshift \citep{ra10,luo10}}
\end{itemize}

\begin{deluxetable*}{rrrcrrccccccc}
\tabletypesize{\scriptsize}
\tablecaption{Optical and near-infrared counterparts to X-ray sources in the \hbox{E-CDF-S} \label{opt_catalog}}
\tablewidth{0pt}
\tablehead{
XID  &  \multicolumn{1}{c}{RA} &  \multicolumn{1}{c}{Dec} &  \multicolumn{1}{c}{Catalog}  &  \multicolumn{1}{c}{dist} &  \multicolumn{1}{c}{LR} &  \multicolumn{1}{c}{Rel} &  \multicolumn{1}{c}{R$_{mag}$} &  \multicolumn{1}{c}{z$_{spec}$} & \multicolumn{1}{c}{Quality}&  \multicolumn{1}{c}{Spectroscopic} & \multicolumn{1}{c}{Class}  & \multicolumn{1}{c}{z$_{phot}$}\\
 & \multicolumn{2}{c}{(J2000)}  & &  ($^{\prime\prime}$) &  &  & (\rm AB) & & flag &catalog\\
}
\startdata
1 & 52.79726 & -27.56042 & WFI-R & 1.3 & 3.9 & 0.70 & 21.4 & --- & --- & --- & --- & ---\\ 
2 & 52.80404 & -27.93025 & WFI-R & 0.8 & 11.4 & 0.98 & 23.2 & --- & --- & --- & --- & ---\\ 
3 & 52.80854 & -28.07248 & WFI-R & 0.8 & 1.3 & 0.64 & 23.9 & --- & --- & --- & --- & ---\\ 
4 & 52.80950 & -27.78539 & WFI-R & 0.5 & 17.1 & 0.99 & 22.9 & --- & --- & --- & --- & ---\\ 
5 & 52.81118 & -28.02890 & WFI-R & 0.6 & 9.9 & 0.98 & 23.5 & --- & --- & --- & --- & $ 0.745_{-0.017}^{+0.003}$ \\ 
6 & 52.81231 & -28.00758 & WFI-R & 0.5 & 0.8 & 0.79 & 26.0 & --- & --- & --- & --- & ---\\ 
7 & 52.81267 & -27.92183 & WFI-R & 0.0 & 52.9 & 1.00 & 20.8 & 1.368 & 2 & 7 & BLAGN & -----\\ 
8 & 52.81864 & -27.82785 & WFI-R & 0.1 & 2.3 & 0.92 & 25.3 & --- & --- & --- & --- & $ 2.212_{-0.022}^{+0.025}$ \\ 
9 & 52.82344 & -27.87511 & WFI-R & 3.2 & 3.2 & 0.94 & 18.5 & --- & --- & --- & --- & $ 0.864_{-0.012}^{+0.004}$ \\ 
11 & 52.82757 & -27.63511 & WFI-R & 0.2 & 2.0 & 0.87 & 24.3 & --- & --- & --- & --- & $ 2.257_{-0.043}^{+0.045}$ \\ 
12 & 52.82790 & -27.68926 & WFI-R & 0.1 & 40.4 & 0.99 & 21.6 & --- & --- & --- & --- & $ 1.058_{-0.002}^{+0.002}$ \\ 
13 & 52.83035 & -27.71372 & WFI-R & 1.1 & 18.3 & 0.99 & 20.3 & 0.400 & 2 & 7 & ALG & -----\\ 
14 & 52.83099 & -27.74238 & WFI-R & 0.1 & 24.5 & 0.99 & 22.8 & 1.218 & 2 & 10 & BLAGN & -----\\ 
15 & 52.83217 & -27.78369 & WFI-R & 0.2 & 3.7 & 0.95 & 24.9 & --- & --- & --- & --- & $ 2.049_{-0.012}^{+0.011}$ \\ 
16 & 52.83272 & -27.74502 & WFI-R & 0.5 & 5.7 & 0.97 & 24.5 & 1.033 & 1.5 & 4 & NELG & $ 1.058_{-0.008}^{+0.007} $ \\ 
17 & 52.83290 & -27.94660 & WFI-R & 0.2 & 9.9 & 0.95 & 23.9 & --- & --- & --- & --- & $ 1.635_{-0.010}^{+0.010}$ \\ 
18 & 52.83515 & -27.92499 & WFI-R & 0.1 & 35.2 & 0.99 & 22.1 & 0.760 & 2 & 7 & NELG & -----\\ 
19 & 52.83604 & -27.72266 & WFI-R & 0.2 & 4.0 & 0.95 & 25.0 & --- & --- & --- & --- & $ 1.356_{-0.017}^{+0.028}$ \\ 
20 & 52.83634 & -27.98900 & GEMS & 0.2 & 48.4 & 1.00 & 22.2 & 0.683 & 2 & 10 & NELG & -----\\ 
\enddata
\tablecomments{An abbreviated version of the table is shown.  The full table is provided online with 762 entries.}
\end{deluxetable*}

We list in Table~\ref{object_stats} the statistics regarding the full sample such as those having spectroscopic redshifts and classifications (e.g., BLAGN, NELG, ALG, STAR);  all optical counterparts discussed from this point on have a reliability, as given in Table~\ref{opt_catalog}, greater than 50\%.  We highlight that our VLT and Keck programs have yielded 248 new spectroscopic redshifts thus reaching a completeness of 55\% (422 out of 762 X-ray sources) and 78\% for the optically-bright population ($R_{\rm AB}<24$).

\begin{deluxetable}{lll}
\tabletypesize{\scriptsize}
\tablecaption{Statistics of \hbox{E-CDF-S} X-ray point sources\label{object_stats}}
\tablewidth{0pt}
\tablehead{
\colhead{Type}&\colhead{ALL}&\colhead{$R_{mag}<24$}}

\startdata
X-ray&762&397\\
Spec + phot $z$&608 (80\%)&372 (94\%)\\
Spec $z$: Flag$>$0&422 (55\%)&310 (78\%)\\
~~~~~~~~~~=2.5&278&236\\
~~~~~~~~~~=2&42&33\\
~~~~~~~~~~=1.5~~~~~~~&26&14\\
~~~~~~~~~~=1~~~~&45&17\\
Spec $z$; New VLT&101&57\\
~~~~~~~~~~New Keck&147&123\\
BLAGN&123&90\\
NELG&188&137\\
ALG&48&37\\
STAR&14&14\\
\enddata
\end{deluxetable}

\subsection{Photometric redshifts}

\label{photoz}

We provide in Table~\ref{opt_catalog} and use through this work new photometric redshifts \citep{luo10,ra10} for 274 X-ray sources in the \hbox{E-CDF-S} that lack secure ($flag<2$) spectroscopic redshifts.  Briefly, these photometric estimates are primarily based on optical photometry in 17 bands, \citep{wo04,wo08}, the GOODS-S MUSIC survey \citep{gr06}, and new data at UV (e.g., from GALEX\footnote{See \url{http://galex.stsci.edu/GR4/}}), optical \citep[e.g., from MUSYC;][]{gaw06}, near-infrared \citep[MUSYC $J$, $H$, and $K_s$;][]{taylor09}, and mid-infrared \citep[e.g., from SIMPLE;][]{dam09} wavelengths that greatly aid in the derivation of photometric redshifts of fainter sources ($m_R \gtrsim 24$) and those at higher redshift ($z\gtrsim 1.4$).  Based on these data sets, we construct a spectral energy distribution (SED) in up to 42 bands for each optically selected source and then utilize the publically-available Zurich Extragalactic Bayesian Redshift Analyzer \citep[ZEBRA;][]{fe06} code to derive photometric redshifts based on 259 galaxy templates constructed from PEGASE stellar population synthesis models \citep[for details, see ][]{gr06}.  We further include a set of hybrid templates (galaxy+AGN) and 10 empirical AGN templates as provided by \citet{po07}.  The ZEBRA code enables us to correct for systematics in the photometry by modifying the input templates to better represent the observed SEDs of the 2651 sources with spectroscopic redshifts.  A maximum likelihood technique is then used to identify the most probable redshift.  We note that photometric redshifts are likely to be more accurate in the central CDF-S region \citep{luo10} due to the deeper data as compared to the extended region \citep{ra10}.  We refer the reader to \citet{luo10} and \citet{ra10} for specific details on the procedure and a detailed quantitative analysis of their reliability especially with respect to the X-ray selected AGN population.


Photometric redshifts can also be used to check the validity of spectroscopic redshifts especially relevant for the faintest sources having considerable uncertainty associated with their redshift determination mainly due to low signal-to-noise spectra.  Therefore, an assessment of the difference between spectroscopic and photometric redshifts is carried out; the spectroscopic redshift quality flag is increased by 0.5 if there is good agreement between the two ($\Delta z (1+z)^{-1}< 0.12$); this cutoff ($2\sigma_{NMAD}$) is based on blind tests \citep[see Figure 12b in][]{luo10}.  Therefore, a low confidence redshift that agrees with its photometric redshifts will have a quality flag 1.5 and likely be correct.    We find that 49\% for the lower reliability sources ($flag=1$) have spectroscopic and photometric redshifts in agreement.  Identifications, based on a single emission line, have a comparable success rate to those determined by the faint continuum as detected in low signal-to-noise spectra.  We can also check the consistency for the secure spectroscopic redshifts ($flag =2$) and find that 79\% are in agreement but caution the reader that the photometric estimates for these sources with reliable redshifts are not independent measurements since their spectroscopic redshifts were used as a training set for the estimate of photometric redshifts.  We provide a bit of caution that these success rates do not necessarily mean that the spectroscopic redshifts are incorrect since spectra of AGN of these luminosities are known to have complexities due to a blend of AGN and host galaxy emission further complicated by the presence of dust for which may be either associated with the nuclear region in the form of a dusty torus or more extended thus attenuating the stellar light.  We provide statistics, in Table~\ref{object_stats}, regarding the fraction of the E-CDF-S X-ray sources with redshift estimates.  Remarkably, we have attained a completeness of 80\% based on both spectroscopic and photometric redshifts that reaches 94\% for those with $R_{\rm AB}<24$.  

\section{Results}
\label{ecdfs}
\subsection{Demographics of the X-ray point-source population}

\label{demographics}

Overall, we find similar optical source populations to previous identification programs in the $Chandra$ Deep Fields \citep[e.g., ][]{ba03,sz04}.  With our deeper exposures, we are able to extend effectively the accessible optical luminosity and redshift range as illustrated below.  We show in Figure~\ref{fig:fx_opt} the distribution of broad-band (0.5--8.0 keV) X-ray flux, as given in \citet{le05}, versus optical magnitude (R) to illustrate the parameter space spanned by the various populations.  Sources are marked by their method of identification, either spectroscopic or photometric, and object type if available based on their optical spectra.  As shown by the aforementioned identification programs, the BLAGN follow the trend of $0\lesssim log(f_X/f_R)\footnote{Ratio between broad-band X-ray flux and optical magnitude is determined by the equation $log(f_{\rm{X}}/f_{R})=log(f_{\rm X})+0.4R+5.69$} \lesssim 1$ with significant scatter.  The NELG are more numerous than the BLAGN, span a wider range in the $f_X-R$ plane, and are the dominant population at $f_X/f_R>10$ \citep{fiore03}.  A significant amount of the dispersion in this population comes from the fact that most NELGs have significant absorbing columns; this is evident by the tighter relationship between X-ray flux and optical magnitude when considering the hard (2--8 keV) band \citep[e.g., ][]{ba03,si05,eck06}.  We further find that $64\%$ of the X-ray bright, optically faint ($f_X/f_R>10$) NELGs are AGNs of moderate-luminosity ($10^{42}\lesssim L_{\rm X} \lesssim 10^{44} ~{\rm erg}~{\rm s}^{-1}$) at $0.5\lesssim z \lesssim1.5$; type 2 QSOs, defined by their lack of broad-emission lines (Section~\ref{class}) and having $L_{0.5-8.0~{\rm keV}}>10^{44}$ erg s$^{-1}$, comprise 30\% of sources having $f_X/f_R>10$ and $R>24$ (7 of 24).  At faint flux levels ($f_{0.5-8.0~keV}\lesssim 5\times10^{-15}$ erg cm$^{-2}$ s$^{-1}$), a significant number of ALGs are evident; they span a wide range of optical magnitude (See Sec.~\ref{abs} for further details).  Based on their X-ray luminosities ($L_{\rm X} \gtrsim 10^{41} ~{\rm erg}~{\rm s}^{-1}$), their X-ray emission is likely powered by an AGN in most cases (See Figure~\ref{fig:lx_z}).  In addition, there are also a handful of stellar sources that have been identified having faint X-ray fluxes and bright optical magnitudes ($15<R<20$).

\begin{figure}
\epsscale{1.15}
\plotone{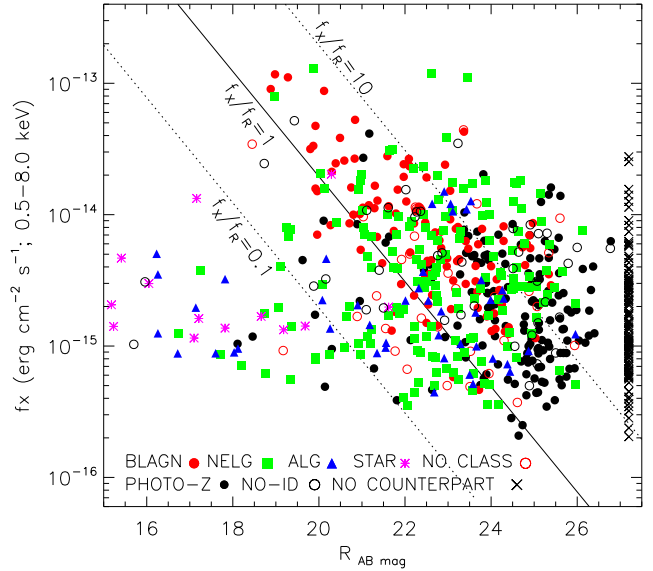}

\caption{Broad-band (0.5--8.0 keV) X-ray flux versus optical magnitude
($R_{AB}$).  X-ray sources with identification through optical
spectroscopy are marked in color with the symbol type denoting the
object classification (BLAGN, NELG, ALG STAR; see section~\ref{class}
for further details).  Sources with photometric redshifts are marked
by filled black circles.  Open
red circles are sources having spectroscopic redshifts but lack
classification.  X-ray sources with no optical counterparts are
arbitrarily placed at R=27 (crosses).  Lines of constant
X-ray-to-optical flux ratio (0.1,1,10) are given by Equation 1 of
\citet[][]{sz04}.}
\label{fig:fx_opt}
\end{figure}

Many of the optical counterparts lacking secure spectroscopic identification (filled black circles) are fainter than $R\sim24$.  Although, as demonstrated in Figure~\ref{complete_type} and~\ref{fig:fx_opt}, our VLT and Keck observations have enabled the identification of a significant number (56 with $flag \ge 1$; 36 with $flag \ge 1.5$) of optically faint ($R>24$) counterparts by their emission lines, thus classified primarily as NELGs, although a fair number do exhibit faint broad lines (See Figure~\ref{fig:spectra_desert}$b-c$ for some examples) that are further discussed below.    Finally, we highlight that there is a significant number of X-ray sources having $R\gtrsim 26$ that can further be characterized with deeper near-infrared imaging \citep{cardamone08,brusa09} that is beyond the scope of the current work; although, we do confirm that all of the brightest X-ray sources ($f_{\rm 0.5-8~keV}>10^{-14}$ erg s$^{-1}$ cm$^{-2}$) without optical counterparts are clearly associated with bright, near-infrared detections.

We further remark that many of the narrow emission lines (e.g., [OIII]$\lambda$5007, H$\beta$, H$\alpha$, [NII]) usually present in X-ray selected NELGs are likely powered by photoionization by the central AGN especially for those having $L_{\rm X}>10^{42}$ erg s$^{-1}$ as demonstrated by COSMOS \citep{bongiorno10} and ChaMP \citep{constantin09}.  Even so, a significant fraction ($\sim10-60\%$) of the [OII]$\lambda$3727 emission-line flux has been attributed to star formation even for the X-ray luminous cases \citep{si09b}.  Such detailed emission-line diagnostics of AGNs in deep X-ray surveys require further spectroscopy in the near-infrared given their typical redshifts ($z\sim1$).

The effectiveness of probing moderate-luminosity AGN ($L_X\sim10^{43-44}$ erg s$^{-1}$) at $z>1.5$ is demonstrated by the X-ray luminosity distribution as a function of redshift shown in Figure~\ref{fig:lx_z}.  The X-ray luminosity has been estimated from the broad-band flux given in \citet{le05} with a $k$-correction based
on a power law spectral energy distribution ($l_E \propto E^{1-\Gamma}$; $\Gamma=1.9$; e.g., \citealt{si05,ma07}).  No correction for intrinsic absorption has been applied thus these luminosities are likely to be underestimated.  The solid line denotes the evolution of the characteristic break ($L_{\star}$) in the luminosity function of AGN, based on the pure luminosity evolution model (Eq. 21) from \citet{aird10}, that illustrates the effectiveness of surveys, reaching depths similar to the \hbox{E-CDF-S}, to detect AGN contributing to the faint end slope of the XLF up to $z\sim3$ \citep[e.g.,][]{ba05, aird08}.  We are now able to fill in the region of $L-z$ space ($L_X<10^{44}$ erg s$^{-1}$; $1.5\lesssim z \lesssim 2.5$) previously lacking identification with objects that have weak emission lines (either broad or narrow).  Our deep VLT exposures with the LR-blue grism effectively enable identification of X-ray sources without strong optical signs of an active nucleus, thus mitigating the strong bias evident in most spectroscopic surveys (i.e., the 'redshift desert'; \citet[See Figure 6 of][]{br05}).

\begin{figure}
\includegraphics[scale=0.55]{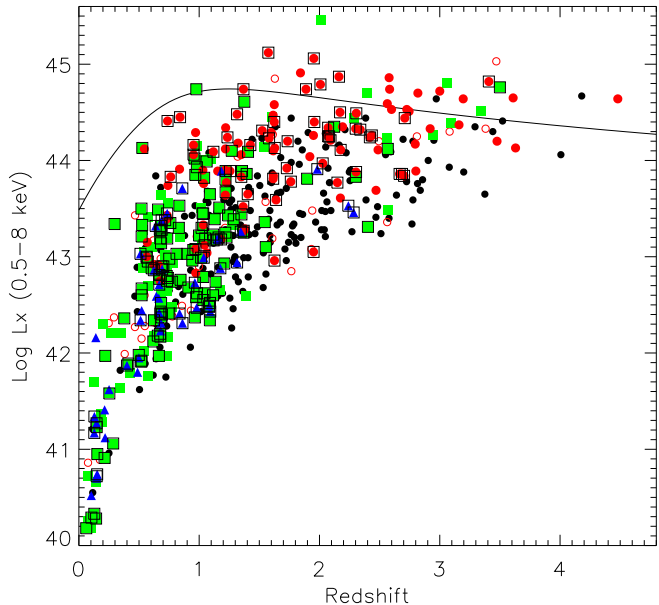}

\caption{Luminosity--redshift distribution of X-ray sources.  Symbols are the same as described in Figure~\ref{fig:fx_opt} with the exception of the open squares that indicate those with new spectroscopic redshifts from our own VLT and Keck programs.  For illustrative purposes, we plot the evolution of the break in the X-ray luminosity function based on analytic model of \citet{aird10}.}

\label{fig:lx_z}
\end{figure}

In Figure~\ref{fig:spectra_desert}, we show optical spectra of four moderate-luminosity AGN that we have identified at redshifts for which wavelength coverage below 5000 ${\rm \AA}$ facilitates object classification.  In the first three panels ($a$: \#162; $b$: \#357; $c$: \#484), the spectra are of very faint counterparts ($R=24.8-26$) and mainly show evidence of faint emission lines clearly visible in the 2D images.  The first spectrum (\#162) is of a counterpart to an X-ray source that has been identified by \citet{don10}, designated as IRBG1, to have the highest 24 $\micron$-to-optical flux ratio ($f_{24}/f_R=3961$) among $Spitzer$ sources detected in deep $Chandra$ fields.  The spectrum here was obtained from our deep 9 hour exposure of the central CDF-S region with the VLT/VIMOS.  The spectroscopic redshift is based not only on the emission lines of CIV and CIII] but interstellar absorption lines of AlII and FeII associated with the host galaxy.  We classify this object as a NELG due to the width of the CIV line to be $\sim$1200 km s$^{-1}$ that supports the conclusion of \citet{don10} that the IR emission from this object is likely to be due to an obscured AGN.  In the second case (panel $b$), the redshift is based on a significant detection of an emission line at 4487 ${\rm \AA}$ that we identify as CIV that may be further supported by a possible detection of CIII] that is uncertain due to a nearby night sky emission feature thus the quality flag is set appropriately.  In this case, the spectroscopic and photometric redshifts are not in agreement although the photometric estimate does indicate that the X-ray source is at high redshift ($z_{phot}=1.154$).  In panel $c$, we have high confidence in the spectroscopic redshift ($z=2.403$) based on a strong Ly$\alpha$ emission line, and CIV emission.  This identification is notable given its faint magnitude ($R=25.96$).  Optical counterparts in panels $b$ and $c$ are classified as type 2 AGNs based on their narrow line widths.  Lastly, we show the spectrum associated with ECDFS \#667 that is optically brighter than the previous three cases thus resulting in a highly secure redshift given by the strong CIV emission, and weaker CIII] lines that are both broad in line width and typical of optically-selected QSOs.

\begin{figure*}
\hspace{2.6cm}
\includegraphics[scale=0.65]{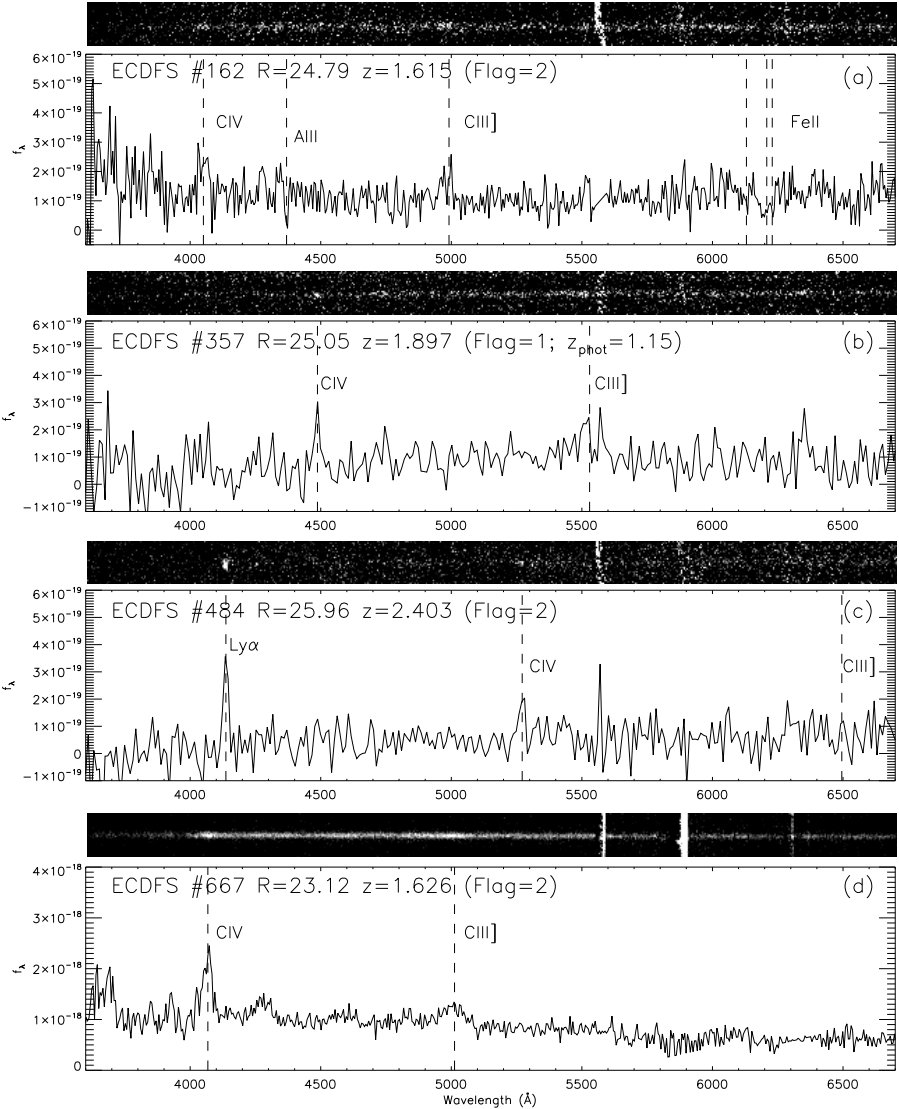}
\caption{Optical spectra from the VLT/VIMOS of four moderate-luminosity ($log~L_X\sim43.5$) AGN at $1.5<z<2.5$.  The units are flux density (erg cm$^{-2}$ s$^{-1}$ ${\rm \AA}^{-1}$).  Spectra in panels $b$ and $c$ have been lightly smoothed with a boxcar of 2 pixels.  The quality of the assigned spectroscopic redshift is given in each case along with its photometric redshift for the one case (panel $b$) with lower confidence.  Each one-dimensional spectrum is accompanied by its 2D image.}
\label{fig:spectra_desert}
\end{figure*}

To conclude here, we highlight that it is important to determine how the more complete sampling of the $L_{\rm X}-z$ plane (Figure~\ref{fig:lx_z})  impacts recent determinations of the AGN XLF particularly the faint-end slope at $1.5 \lesssim z \lesssim 3$.  Such accurate assessments of the XLF are crucial to better constrain our understanding of the evolution of the black-hole mass function \citep[e.g.,][]{ta06,me08}.  For example, \citet{ye09} claim that they have a high level of spectroscopic completeness with respect to the identification of BLAGNs thus the flatness of the observed faint-end slope is robust.  We assert that this may not be the case since many of the optically-faint X-ray sources identified herein do have broad emission lines thus the faint-end slope may actually be steeper than observed in spectroscopically-selected samples.  Recent evidence does suggest that incompleteness with respect to accurate redshift determination is likely to have a significant impact on the measurement of the faint-end slope at $z>1$ \citep{aird10}.  Therefore, it is imperative to place stringent constraints on the faint-end of the luminosity distribution of AGNs with a sample that has secure redshifts down to the faintest optical limits feasible, such as that presented here.   Given the complexities in properly dealing with multi-band selection and incompleteness, we reserve such an analysis to a separate study now underway.  


\subsubsection{Type 2 QSOs}

X-ray surveys with $Chandra$ \citep[e.g.,][]{st02,no02,ba03,alex08} and XMM-$Newton$ \citep{dw06,ma07,vignali09} are now amassing samples of obscured (i.e., type 2) QSOs, a previously elusive population thought to be prevalent based on unification models and the spectral shape of the CXRB.  These sources are typically characterized by their high luminosity in the X-ray band ($L_X>10^{44}$ erg s$^{-1}$) and significant obscuration in both the optical, thus resulting in a lack of broad ($FWHM>2000$ km s$^{-1}$) emission lines, and X-ray, with absorbing column densities greater than 10$^{22}$ cm$^{-2}$.  Based upon their detection rate in X-ray surveys, this population is, at most, equivalent in size to the type 1 QSO population  \citep{tr06,gi07,ha08,alex08}, and unlikely to match the proportions (i.e., obscured-to-unobscured ratio) seen in the less-luminous AGNs.  Although, deeply embedded (i.e., Compton-thick) QSOs continue to be found \citep{alex08} that may elevate the known space density of galaxies having significant obscured accretion \citep[e.g.,][]{pol06,to06,daddi07,fiore09,tr09b}.  In any case, these type 2 QSOs likely represent an important piece to the puzzle in order to understand the mechanisms driving the evolution of SMBHs in general.  For example, \citet{fa08,fa09} argue that the dearth of obscured ($log~N_H\gtrsim 21.5$) AGNs accreting close to their Eddington rate is due to a feedback mechanism such as radiation pressure that effectively clears out the obscuring medium.  Therefore, an improvement in the statistics is warranted especially in a survey field of remarkable multi-wavelength coverage to determine accurate black hole masses and accretion rates.

\begin{figure*}
\epsscale{1.0}
\plottwo{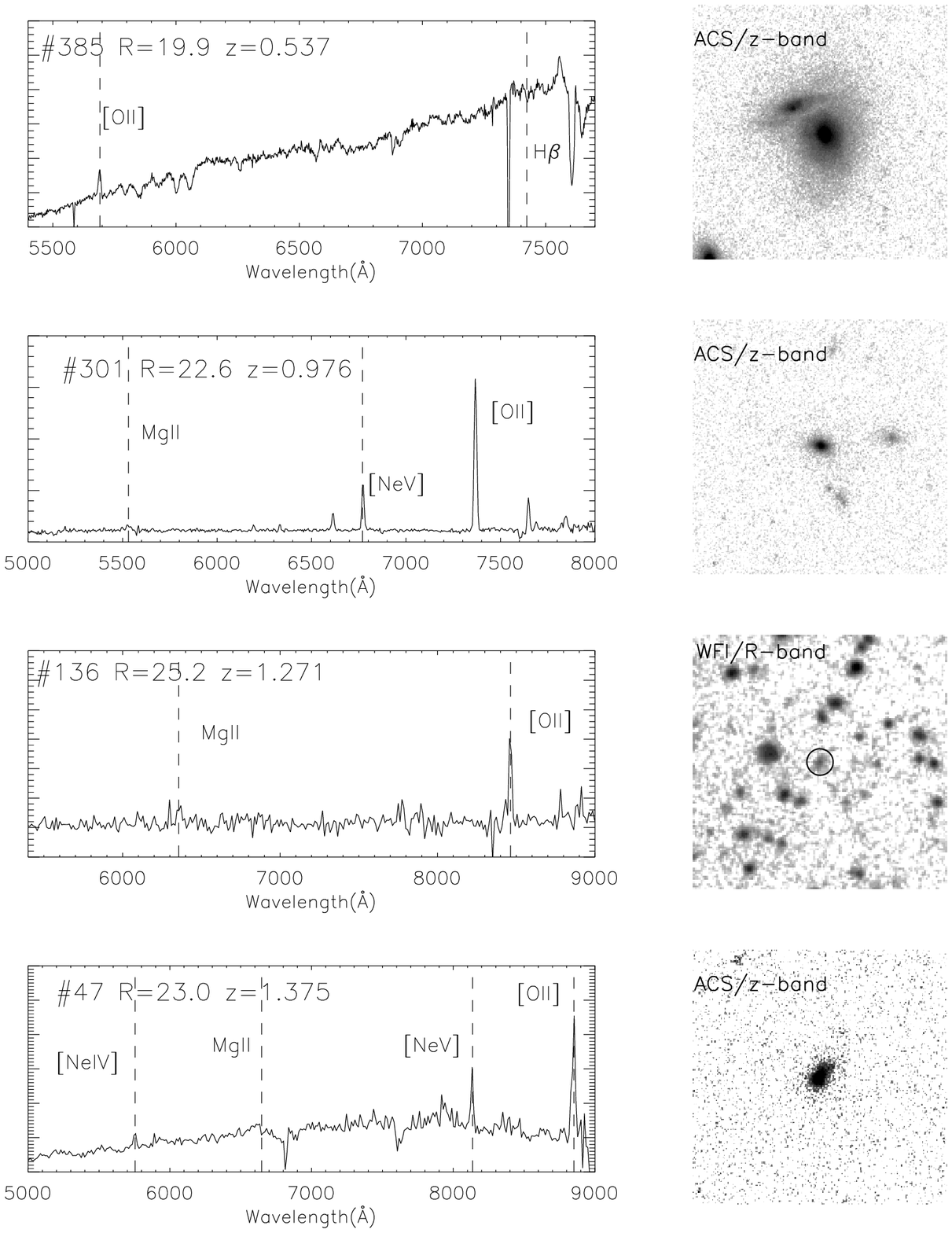}{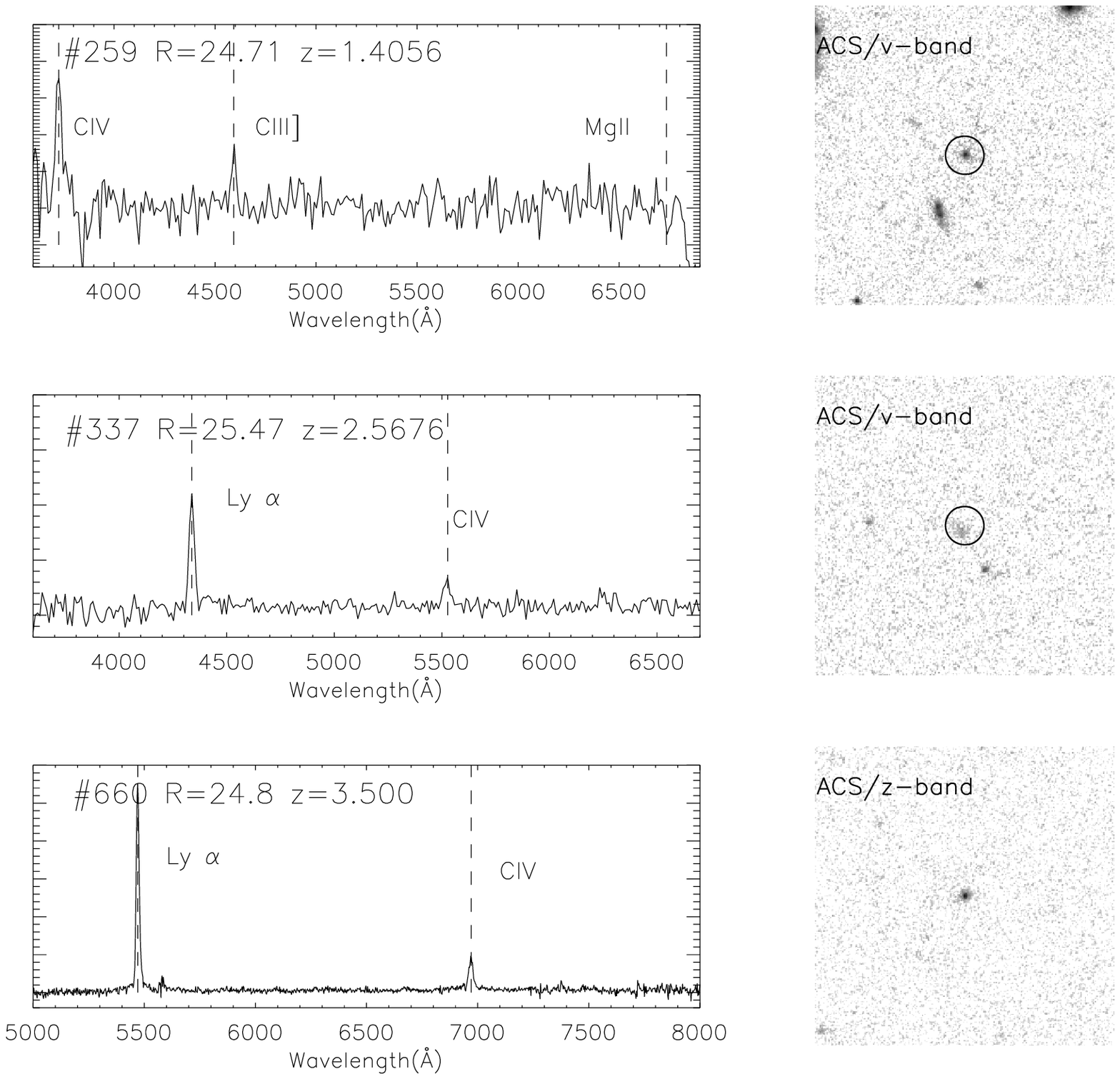}

\caption{Newly-identified Type 2 QSO candidates in the \hbox{E-CDF-S}.  The
\citet{le05} catalog number, optical magnitude and redshift are
labelled in each panel.  Black circles are shown in some image cutouts to aid in the identification of the target.}

\label{fig:type2qso} \end{figure*}

To gauge the improvement in the numbers of type 2 QSOs in our newly compiled sample, we find that there are 17 X-ray sources in the broad-band selected sample that have a secure spectroscopic redshift ($Flag \ge 1.5$), $L_{0.5-8.0 {\rm keV}}>10^{44}$ erg s$^{-1}$, and an optical spectrum with only narrow emission lines.  If we relax the criteria on the quality flag for the redshift ($Flag \ge 1$), the sample increases substantially to 33.  We have identified 7 out of the 17 high-quality identifications through our spectroscopic program.  In Figure~\ref{fig:type2qso}, we display our optical spectra and image cutouts of this group having redshifts spanning a wide range $ 0.5 < z < 3.5$.  Two of the lower redshift QSOs (\#47, \#301) have [NeIV] and [NeV] emission lines fully consistent with ionization from a non-stellar continuum \citep{gilli10}.  The highest redshift QSOs (\#337, \#660) are very similar to those previously identified cases mentioned above \citep{st02, no02}.  We have performed X-ray spectral fits (Bauer et al. in preparation) to check whether these sources have high levels of absorption, as expected given the significant amount of extinction needed to hide the broad line region.  For the five QSOs that have sufficient X-ray counts ($>50$) in the 0.5--8.0 keV band to perform a fit with a power law model having two free parameters ($\Gamma$, $N_{H}$), we have measured X-ray absorbing column densities in the range of $8.25-24.0\times10^{22}$ cm$^{-2}$ thus indicative of values typical for X-ray selected type 2 QSOs \citep[e.g.,][]{to06,ma07}.  Further support of their 'type 2' nature requires deep near-infrared spectroscopy to rule out a broad component to the H$\alpha$ and H$\beta$ emission lines \citep{akiyama02}.  We conclude by emphasizing that the optical magnitudes of these X-ray sources are very faint ($R_{AB} \sim 24-25$) thus further deep spectroscopy, particularly covering the blue end of the optical spectrum, is warranted to make accurate estimates on the obscured fraction of luminous QSOs at high redshifts.

\subsubsection{AGN in unremarkable galaxies at $z>1.5$}
\label{abs}

\begin{figure*}
\epsscale{0.85}
\plotone{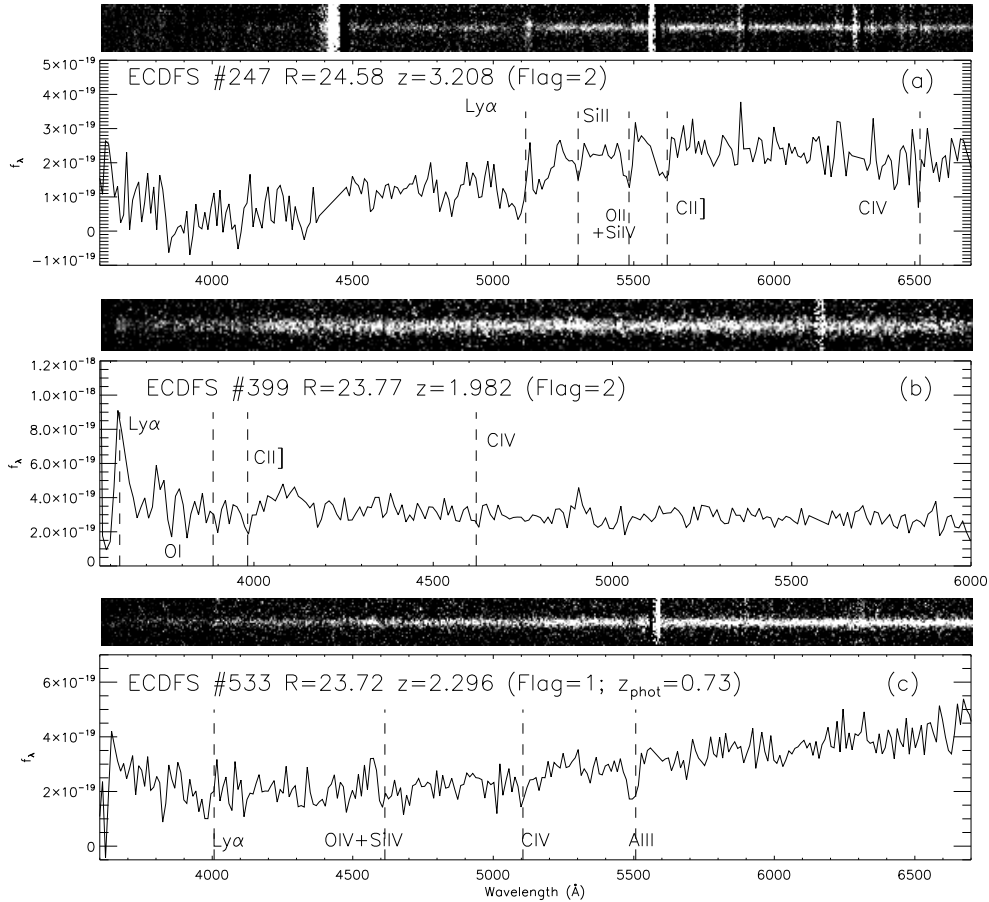}

\caption{Optical spectra of high-redshift ($z\sim2-3$) AGN in normal galaxies.  The units are flux density (erg cm$^{-2}$ s$^{-1}$ ${\rm \AA}^{-1}$).  These spectra were acquired with VIMOS using the LRblue grism.  The data are binned by a factor of 2 here thus producing a resolution of 10.6 {\rm \AA} pix$^{-1}$.  Spectroscopic features are labeled with the quality flag on the spectroscopic redshift and photometric redshift estimate given in parenthesis for the case ($c$) having low confidence.  Each 2D image is shown as in the previous figure.}

\label{alg1}
\end{figure*}

For almost three decades, it has been known that X-ray observations are able to identify AGNs that lack optical signs, including narrow emission lines, of an active nucleus \citep[e.g.,][]{elvis81,ma03,se03,cac07}.  Their nature has often been attributed to excess extinction \citep[e.g.,][]{co02,ci07} possibly related to their host galaxies being viewed edge-on \citep{ri06}.  Additional explanations include dilution by the host galaxy \citep[e.g.,][]{mo02,geo05}, unusually weak ionizing continua possibly related to a radiatively-inefficient accretion flow \citep{yuan04}, or beamed emission in the form of a relativistic jet \citep{ca97,worrall99}.  These AGN are typically X-ray faint, thus the $Chandra$ Deep fields are providing significant samples for detailed studies.  To date, most objects in this class have been identified at $z\leq 1.2$ due to the difficulty of identifying galaxies without strong spectral features at higher redshifts.

With our current spectroscopic campaign, we have identified, for the first time, four such systems at $z>1.5$ (mainly due to our deep exposures) with two of them having highly secure spectroscopic redshifts.  Three ALGs are present in the broad-band (0.5--8.0 keV) sample while the other one is only detected in the soft-band (0.5--2.0 keV).  Optical spectra are shown in Figure~\ref{alg1}$a-c$ with accompanying 2D images that further demonstrate the robustness of the continuum and the reliability of absorption lines in these systems.  Their faint optical magnitudes ($R_{\rm AB}\sim24-25$) have likely hampered previous identification programs.  Spectroscopic redshifts are determined by the presence of interstellar absorption lines (e.g., OII+SiIV, CII], CIV, AlII), and the continuum shape as compared to high-redshift galaxy templates that provide further assurance of their redshifts.  

We first highlight the spectroscopic identification of an ALG at \hbox{$z=3.208$} (Figure~\ref{alg1}$a$), the most distant of such galaxies known to date.  This X-ray source is detected by $Chandra$ in the soft-band (0.5--2 keV; ECDFS \#247).  Even though only 6 counts have been detected, the source is reliable given the WAVDETECT threshold of $1\times10^{-6}$, two adjacent pixels each having two detected photons, photon arrival times that are roughly equally spaced, and the X-ray source is clearly coincident with an optical counterpart.  We show the optical spectrum obtained from the 9hr exposure with the VLT.  The 2D spectrum clearly illustrates the existence of absorption lines (Ly$\alpha$, SiII, OII+SiIV, CII]), and a faint narrow Ly$\alpha$ line clearly evident within the absorption trough that appears to be spatially extended ($\sim2.5\arcsec$; 19 kpc).  In a second case (Fig.~\ref{alg1}$b$; ECDFS \#399), we base the spectroscopic redshift on an absorption line at $\sim$4000~{\rm \AA} that is attributed to CII] and further supported by a possible detection of Ly$\alpha$ in emission that falls close to the end of the spectrum.  Interestingly, there is no hint of CIV emission that is a typical signature for AGN activity.  In panel $c$, we show a spectrum of an ALG that has a possible redshift of $z=2.296$ although with considerable uncertainty (flag=1).  The fourth high redshift ALG (not shown) is of similar significance.  We conclude that the X-ray emission, from this class of objects at these high redshifts, is dominated by accretion onto a SMBH based on their high X-ray luminosities ($10^{43}<L_X<10^{44}$ erg s$^{-1}$).

\section{Tracing large-scale structure with AGNs and galaxy groups}
\label{lss}

\subsection{Extended X-ray sources and group/cluster identification}
\label{groups}

\begin{deluxetable*}{llllllll}
\tabletypesize{\scriptsize}
\tablecaption{Identification of extended groups/clusters\label{table:groups}}
\tablewidth{0pt}
\tablehead{
\colhead{Index}&\colhead{XRA}&\colhead{XDEC}&\colhead{ORA}&\colhead{ODEC}&\colhead{Redshift}&\colhead{Type}&\colhead{Telescope}}
\startdata
29&53.07578&-27.78844&53.07515&-27.78851&0.7342&NELG&VLT/VIMOS\\
21&53.19061&-27.68395&53.19500&-27.68726&0.7328&ALG&VLT/VIMOS\\
&&&53.18832&-27.68261&0.7324&ALG&VLT/VIMOS\\
&&&53.18774&-27.68598&0.7328&ALG&VLT/VIMOS\\
&&&53.18543&-27.68917&0.7282&NELG&VLT/VIMOS\\
&&&53.19051&-27.70362&0.7329&NELG&VLT/VIMOS\\
56&53.34735&-27.77114&53.34654&-27.77106&0.8347&NELG&VLT/VIMOS\\
28&53.33588&-27.79903&53.33691&-27.79873&0.1288&ALG&VLT/VIMOS\\
55&53.28957&-27.76900&53.33158&-27.83082&0.6879&ALG&VLT/VIMOS\\
30&52.96078&-27.82475&52.96048&-27.82368&0.6812&NELG&VLT/VIMOS\\
&&&52.95444&-27.82728&0.6820&ALG&VLT/VIMOS\\
&&&52.95737&-27.82756&0.6803&NELG&KECK/DEIMOS\\
31&52.96858&-27.83918&52.96692&-27.84015&0.6792&ALG&KECK/DEIMOS\\
10&52.99811&-27.90738&52.99954&-27.90867&0.7381&ALG&KECK/DEIMOS\\
&&&52.99481&-27.90663&0.7385&ALG&KECK/DEIMOS\\
\enddata
\end{deluxetable*}

The \hbox{E-CDF-S} not only includes unresolved X-ray sources, but also a significant number of extended sources clearly associated with galaxy overdensities (i.e., groups and clusters; \citealt{gi02,ros02,le05}).  The identification of such structures enables us to compare their large-scale distribution with that of AGNs.  Here, we utilize both the $Chandra$ and XMM-$Newton$ observations of the \hbox{E-CDF-S} to construct an extended source catalog (Finoguenov et al. in preparation) and identify potential galaxy members that are then targeted for optical spectroscopy.

\begin{figure*}
\hspace{2cm}
\includegraphics[scale=0.35]{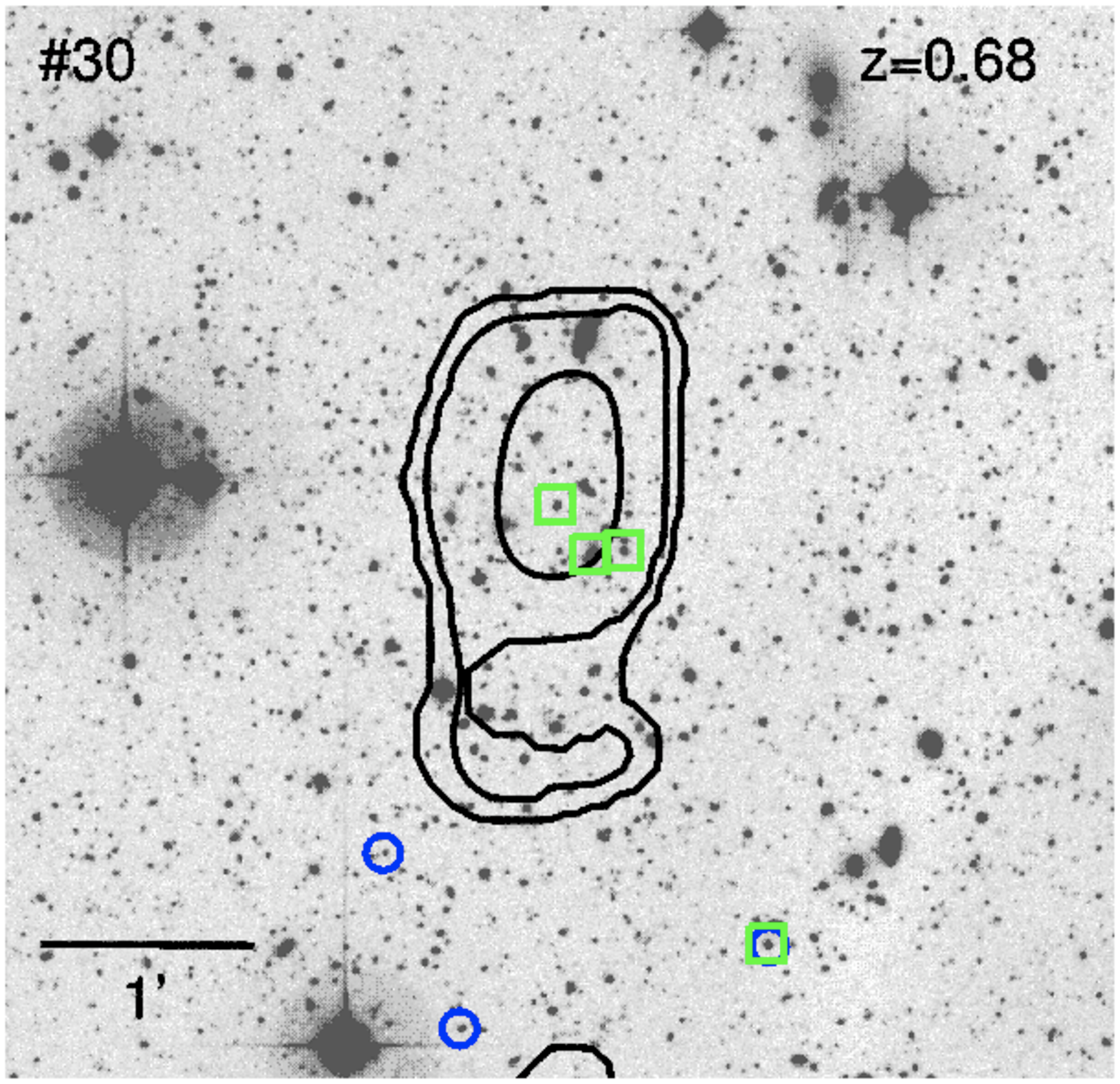}
\includegraphics[scale=0.35]{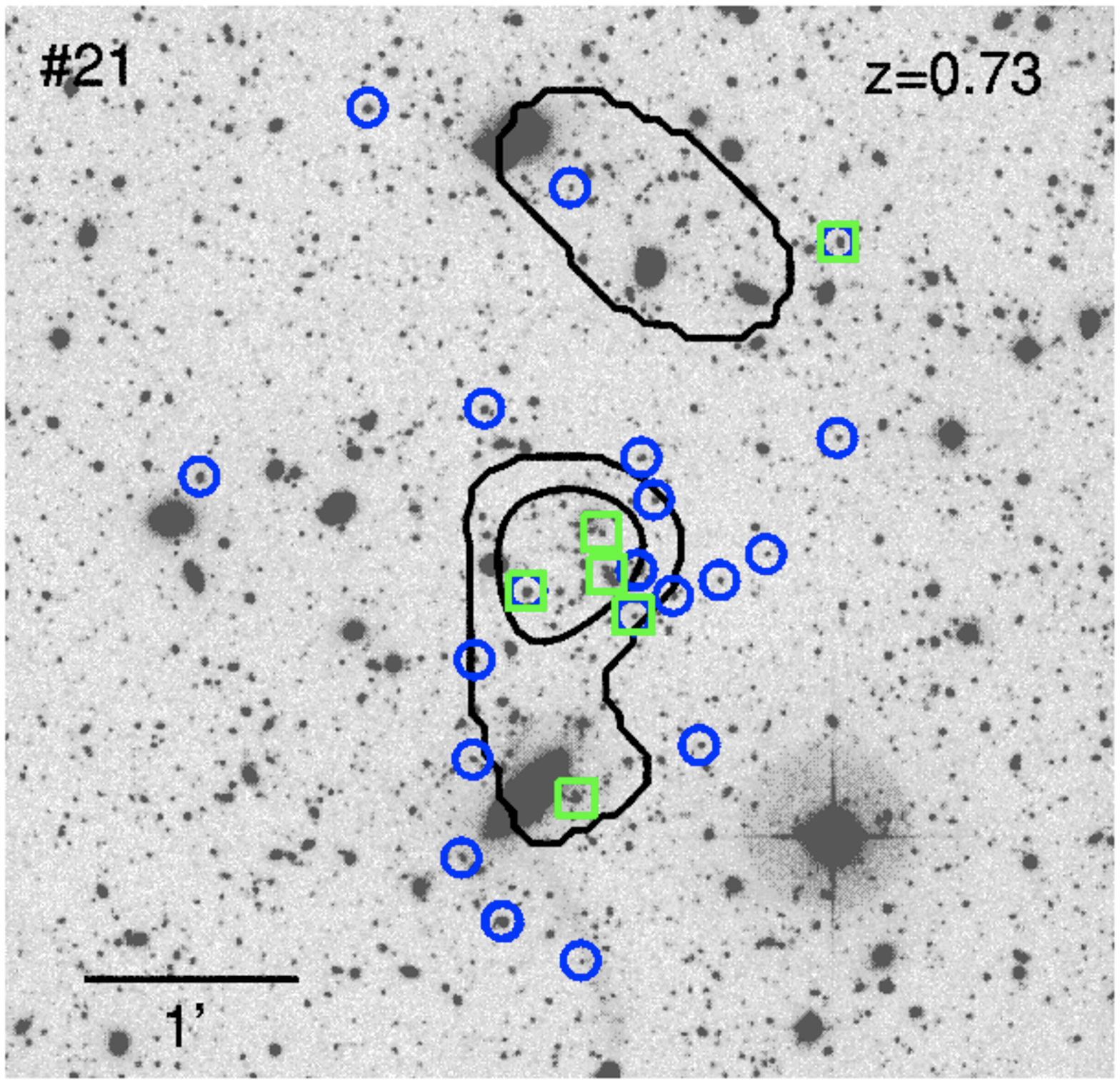}

\caption{Two examples of spectroscopically-identified groups/clusters
(Table~\ref{groups}; \#30, \#21) with extended X-ray emission detected by XMM-$Newton$ and $Chandra$ as shown by the contours overlaid onto the WFI $R$-band image.  Galaxies within a narrow redshift interval ($\Delta z \sim 0.01$) are labelled:  green triangles mark those with spectroscopic redshifts from our program while blue circles are from the GOODS spectroscopic master catalog.}
\label{groups}
\end{figure*}

We briefly describe the XMM-$Newton$ observations and subsequent analysis; we refer the reader to \citet{fino09} for details on the extended source detection method using combined $Chandra$ and XMM-$Newton$ data.  The exposure time of the XMM-$Newton$ observation is 541 ks; a final exposure of 309 ks was left after filtering events for periods of low background as described in \citet{zhang04}.  We follow the detailed background removal procedure of \citet{fi07} and monitor the appearance of atypical background pattern in MOS chips as identified in \citet{snow08}.  The next step entails the removal of point sources from our background-subtracted and vignetting-corrected images.  We implement a source detection algorithm employing wavelets at the scales of 8 and 16 arcseconds.  Subsequently, we restore the spatial distribution of the source flux based on knowledge of the PSF.  After subtracting a model image, we proceed with the analysis of the extended emission.  The final catalog of extended objects contains 52 sources, many of which have previously been identified \citep{gi02,le05}.  Newly detected sources are then compared with photometric galaxy catalogs with most having a clear association to an overdensity of galaxies.

We have targeted for spectroscopic observations a subset of galaxies that are potentially associated with these galaxy groups.  In Table~\ref{table:groups}, we list galaxies that have a secure ($flag=2$) spectroscopic redshift ($0.6\lesssim z \lesssim 0.8$) thus providing a likely distance to eight extended X-ray sources given their agreement with photometric redshift estimates of associated group galaxies.  At the time of the spectroscopic observations, no spectroscopic confirmation had been obtained for these groups.  With the recent spectroscopic campaign from GOODS \citep{balestra10},  six of the eight groups have further confirmation of their redshift.  In Figure~\ref{groups}, we show two examples, with the galaxies that are used for the redshift determination labelled.  The estimated X-ray luminosities ($L_{0.5-2.0~{\rm keV}}\sim10^{41}-10^{42}$ erg s$^{-1}$) indicate that these new systems are low-mass galaxy groups ($M_{halo}\sim10^{13}$ M$_{\sun}$; Finoguenov et al., in preparation) thus improving upon such samples at high redshift \citep[see][for recent progress on the $L_X-M_{halo}$ relation]{le10}.  The statistical properties of the extended X-ray sources revealed by these observations will be presented in Finoguenov et al. (in preparation).  Below, we utilize our new galaxy group catalog to map further the large-scale features in the CDF-S that are typically traced by AGNs.

\subsection{Physical extent of large-scale structures}

Deep X-ray surveys provide a unique opportunity to measure the clustering strength of AGNs on $\sim$10 Mpc scales by utilizing the prevalence of moderate-luminosity ($logL_X$$\sim$43) AGNs.  Exceptional structures have been observed in the 1 Ms CDF-S at z=0.67 and z=0.73 \citep{gi03} that are narrow in redshift space ($\delta$$z \lesssim 0.02$) and contain about 1/3 of the identified X-ray sources.  These sheets, equally populated with X-ray sources, are also clearly evident in the galaxy distribution \citep[K20;][]{ci02} with the $z$=0.73 spike having about twice as many galaxies as the one at $z$=0.67, suggesting that AGN activity may be sensitive to large-scale environment \citep{si08a,lehmer09}.  Although, the dependency of AGN activity on host-galaxy stellar mass \citep[e.g.,][]{ka04,si09a,brusa09,xue10} can mimic signs of an environmental effect.  Similar redshift features have been found in the CDF-N, though far less prominent (1/8 of the X-ray sources) and are evidence of the large cosmic variance between the deep fields: the correlation length in the CDF-S ($r_0$=8.6$\pm$1.2 Mpc) is twice that in the CDF-N \citep[$r_0$=4.2$\pm$0.4 Mpc;][]{gi05}.  The AGNs in these two narrow redshift intervals are spread across the CDF-S field of view (17$\arcmin$x17$\arcmin$) corresponding to a linear size of $>7$ Mpc; the $z=0.73$ structure is more compact \citep{ad05,sal09}. These large-scale structures also include galaxy groups/clusters with extended X-ray emission, thus providing a tracer of the cosmic web which we further map here with the extended area coverage of the field.  Furthermore, we investigate the existence of higher redshift structures identified including those at $z$=1.62 and 2.57, which have been typically traced with only 4-5 objects \citep{gi03}.

\begin{figure}
\epsscale{1.1}
\plotone{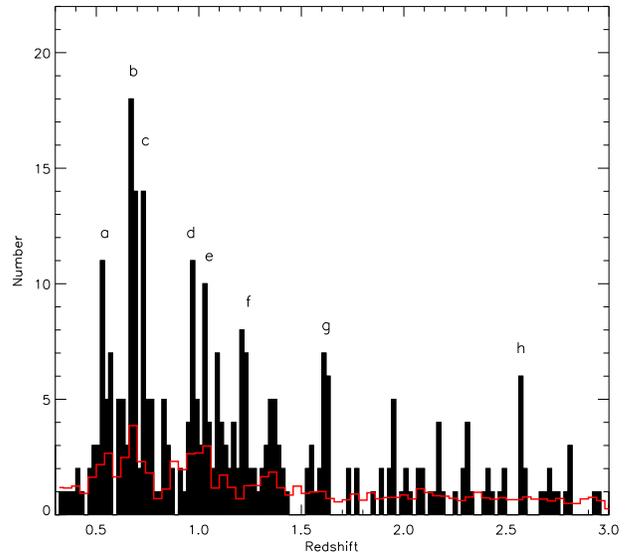}

\caption{Spectroscopic redshift distribution of \hbox{E-CDF-S} X-ray sources up to $z=3$.  Data have been binned with an interval of $\Delta z=0.02$.  For comparison, we plot the distribution of galaxies with photometric redshifts from GOODS-MUSIC (red histogram).  Discrete features, based on the spectroscopic sample and described in the text, are labeled by letter.}

\label{zdistr}
\end{figure}

We show a histogram of the redshift distribution in Figure~\ref{zdistr} to determine whether the highly structured features within the central 1 Ms CDF-S \citep{gi03} are evident on larger scales.  We label the spectroscopic features discussed below with a letter designation ($a-h$).  Overall, the redshift distribution is clearly more evenly populated as compared to that shown in Figure 1 of \citet[][]{gi03}, thus effectively reducing the level of cosmic variance seen in the 1 Ms CDF-S.  The well-known dominant spikes at $z=0.67$ (b) and $z=0.73$ (c) persist and extend beyond the central region (see below for further discussion on their physical extent).  At lower redshifts, we confirm the significant overdensity at $z=0.53$ (a), having 11 AGN in our sample, as reported by \citet{tr09a};  this result agrees with the galaxy distribution from the K20 survey.  We note that the redshift distribution below $z<0.5$ is dissimilar to that presented in \citet{tr09a} with less pronounced features in our sample.  At higher redshifts ($z\sim1$), we not only confirm the previously-reported redshift spike at $z=1.04$ (e) with better statistics (10 AGN) but find an equally populated overdensity at $z=0.97$ (d), not seen in the K20 galaxy distribution but detected using a detailed analysis of the GOODS-MUSIC photometric catalog \citep{sal09}.  At even higher redshifts,  we confirm the enhancements previously seen at $z=1.62$ (g) and 2.57 (h), and identify a new feature at $z=1.22$ (f) that is not spatially concentrated on the sky.  As a final note, we show the distribution of galaxies with photometric redshifts from GOODS-MUSIC (Fig.~\ref{zdistr}) and find that most of the structures discussed here are smoothed out; the only two identified features are enhancements at $z=0.67$ and $z\sim1$ that are actually composed of two distinct structures as detailed above.

\subsubsection{Structures at $z\sim0.7$}

\begin{figure*}
\hspace{2.5cm}
\includegraphics[scale=0.7]{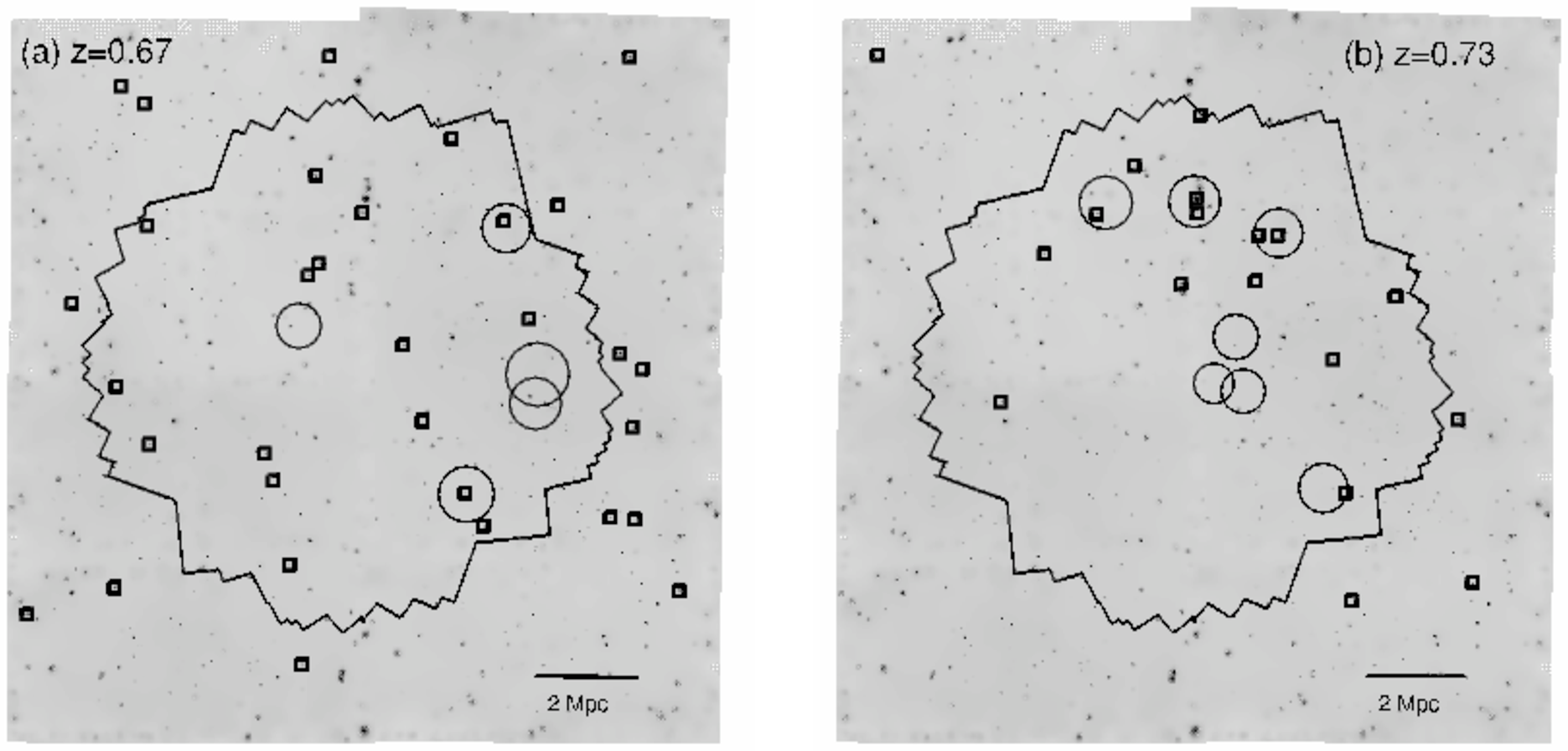}
\vspace{3cm}
\vskip -1.5cm
\hskip 5cm
\includegraphics[scale=0.55]{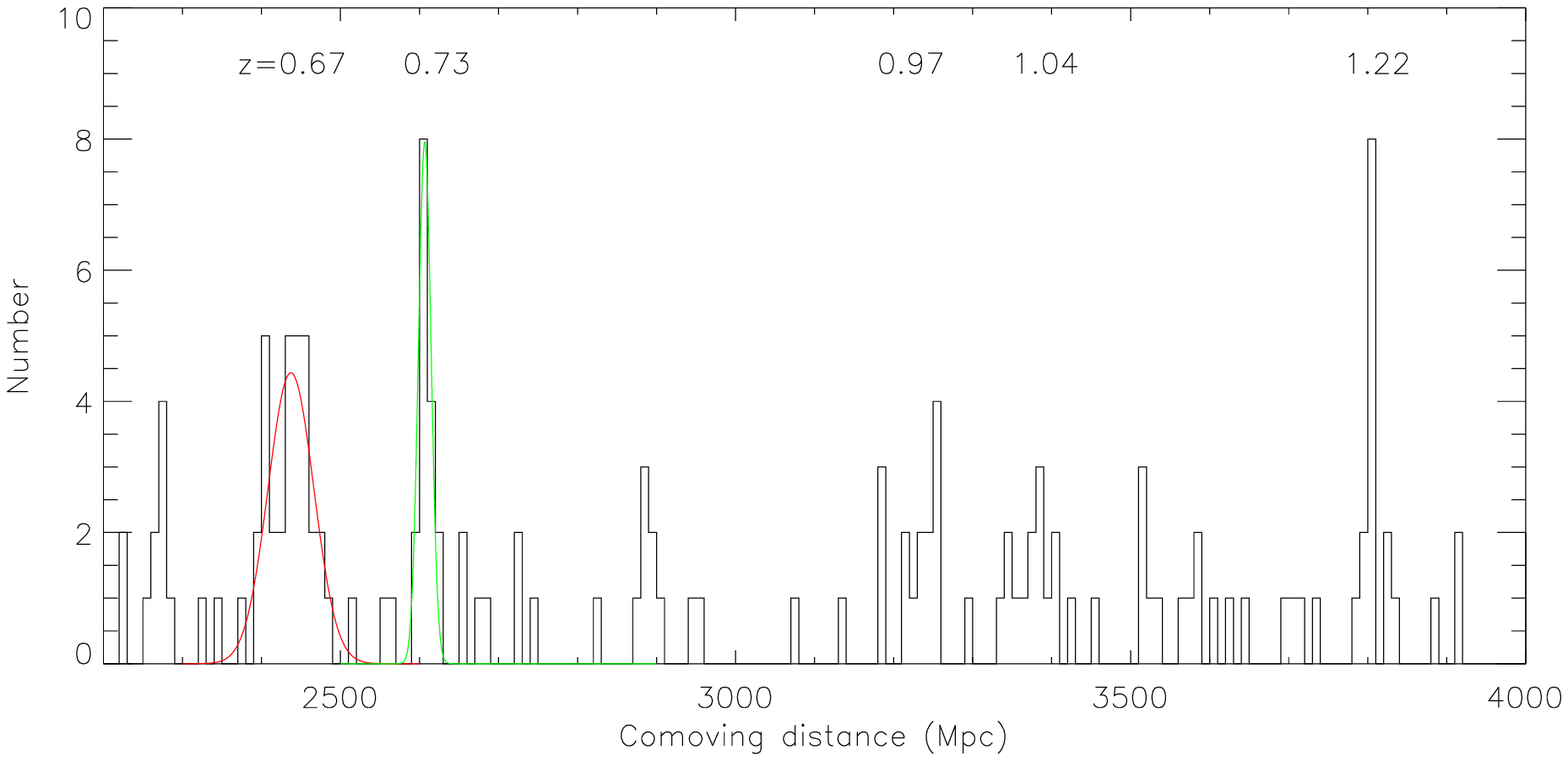}
\vskip 1cm
\caption{$Top:$ Spatial distribution of AGNs (squares) having spectroscopic redshifts within the two main redshift spikes: ($a$) $z=0.67$ ($\Delta z=0.04$), ($b$) $z=0.73$ ($\Delta z=0.034$).  Black circles mark the positions of spectroscopically-confirmed galaxy groups associated with extended X-ray emission with their radius indicative of r$_{200}$.  The grey scale image is a smoothed version of the $Chandra$ mosaic of the \hbox{E-CDF-S} . $Bottom:$ Co-moving radial distance over the redshift interval $0.53<z<1.32$ for AGNs with spectroscopic redshifts.  Prominent structures are evident at $z$=0.67, 0.73, 0.97, 1.04, and 1.22.  The well-known features at $z\sim0.7$ are fit by a gaussian function, as shown in color, that have resulting widths (FWHM) of 67.7 (red) and 18.8 (green) Mpc.}
\label{zspikes:spatial}
\end{figure*}

We show in Figure~\ref{zspikes:spatial} the angular extent of spectroscopically-confirmed AGN and galaxy groups falling within the redshift spikes at $z=0.67$ (panel $a$; $\Delta z$=0.04) and $z=0.73$ (panel $b$;  $\Delta z=0.034$).  For the slightly lower redshift spike ($z\sim0.67$), it is evident that a significant number of AGN fall outside the area covered by the 2 Ms field.  Based upon visual inspection, there do appear to be slightly more AGNs and galaxy groups in the southwestern quadrant, although no clear association between AGN and galaxy groups is evident.  Based on this distribution, we estimate that the structure has a physical angular extent of $\gtrsim12$ Mpc.  The width is shown in Figure~\ref{zspikes:spatial} ($bottom$ panel) and has a comoving dimension of 67.7 Mpc based on a Gaussian fit to the redshift distribution; we describe this structure as a thick slab rather than a narrow sheet.

We find that the structure at $z=0.73$ differs in both angular extent and radial size as compared to the structure at $z=0.67$.  The AGNs are not spread over the full field-of-view of the \hbox{E-CDF-S} and likely span a well-defined region (Fig.~\ref{zspikes:spatial}$b$) that is linear in shape with dimensions $\sim10$ Mpc (length) $\times 3$ Mpc (width) and at a position angle of roughly 45 degrees on the sky.  Remarkably, the seven spectroscopically-identified galaxy groups appear to trace similar structure as the AGNs.  We also find that the width, along our line-of-sight, is significantly narrower than the structure at $z=0.67$ (Fig.~\ref{zspikes:spatial}~$bottom$).  We measure a comoving radial extent of 18.8 Mpc.  Therefore, it appears that a narrow filament falls across the field-of-view of the \hbox{E-CDF-S} that provides an effective indication that such structures are conducive to both accretion onto SMBHs and the formation of the potential wells of galaxy groups.

\section{Summary}

Our program to acquire optical spectra and reliable redshifts for $Chandra$ sources in the \hbox{E-CDF-S} is presented.  We utilize multi-slit instruments on both the VLT and Keck to identify the faintest X-ray sources by implementing deep exposures ranging from 2-9 hours.  These observations result in new spectroscopic redshifts for 283 X-ray sources, including a subset belonging to the deeper 2 Ms data; our high identification rate of X-ray sources reaches 80\% with the inclusion of photometric redshifts.  Based on all available catalogs, there now exist over 500 X-ray sources in the full CDF-S area that have a spectroscopic redshift.

We highlight a number of significant and unique contributions that this program brings to the study of the faint X-ray source population. 

\begin{itemize}

\item Improved coverage of the accessible luminosity-redshift plane that includes AGNs, both with and without broad emission lines, spanning the faint end of the luminosity function at $1.5<z<3$. 

\item The secure identification of seven new type 2 QSOs ($0.5 \lesssim z \lesssim 3.5$) in the \hbox{E-CDF-S} field.

\item The use of interstellar absorption (e.g., SiII, OI+SiIV, CII, CIV]) lines to identify an elusive population of optically-faint galaxies ($R_{mag}\sim{25}$) at $z\sim2$; we highlight the highest redshift absorption-line galaxy ($z=3.208$);  this object has no obvious signs of an AGN based on a deep (9hr) optical spectrum.  

\item Improved AGN and galaxy groups samples that further aid our ability to trace large-scale structures evident in the CDF-S.  For example, the previously known redshift spikes ($z=0.67$ and 0.73) do extend beyond the original CDF-S field.  The higher redshift structure ($z=0.73$) appears to be more filamentary.   

\end{itemize}

Further spectroscopic observations with VLT and the Keck are currently being undertaken to increase the level of completeness with respect to the identification of X-ray sources in deep $Chandra$ fields.  In particular, two more deep exposures (9hr each) in the \hbox{E-CDF-S} with the VLT are currently being executed that target the optically faint population.  With this remarkable data set, we expect to further our understanding of the clustering of moderate-luminosity AGN, the luminosity function, and the relation of AGN activity to their larger-scale environments that all shed light on the mechanisms responsible for the growth of SMBHs and the galaxies in which they reside.   

\acknowledgments

The authors recognize support from Michael Cooper for the use of the DEEP2 pipeline that was developed at UC Berkeley with support from NSF grant AST-0071048.  WNB, BL, and YX acknowledge support from the Chandra X-ray Center grant SP8-9003A and NASA ADP grant NNX10AC99G.  GS acknowledges support of the Pol\'anyi Fellowship of NKTH.  DA is funded by the Royal Society and Leverhulme Trust.  MS and GH acknowledge support bei the Leibniz Prize of the Deutsche Forschungsgemeinschaft, DFG (HA 1850/28-1).   AC is supported through ASI-INAF grants  I/023/05/00 and  I/088/06.  PT acknowledges financial contribution from contract ASIÐINAF I/088/06/0.  ``The authors wish to recognize and acknowledge the very significant cultural role and reverence that the summit of Mauna Kea has always had within the indigenous Hawaiian community.  We are most fortunate to have the opportunity to conduct observations from this mountain." 



Facilities: \facility{VLT:Melipal(VIMOS),Keck:II (DEIMOS),CXO (ACIS-I)}

\appendix

\section{Optical identification of additional targets}

\subsection{X-ray sources in the 2 Ms CDF-S}

During our observing campaigns with both the VLT and Keck, we placed additional slits on optical counterparts to X-ray sources in the 1 and 2 Ms catalogs \citep{gi02, luo08} that have fluxes below the limit of the \hbox{E-CDF-S} survey \citep{le05}.  We use the optical counterpart as reported in \citet{gi02} based on VLT/FORS1 images or that from the EIS survey \citep{ar01}.  The optical counterparts to X-ray sources solely detected in the 2 Ms catalog are reported in \citet{luo08} based on a WFI $R$-band image.  In total, we have spectroscopically-identified 49 additional X-ray sources that are listed in Table~\ref{table:2ms_ids}.

\subsection{Counterparts to radio (VLA) sources}

To fully exploit the multiplex capabilities of VLT/VIMOS and KECK/DEIMOS, we have acquired spectra of optical counterparts to radio sources as secondary targets.  The full \hbox{E-CDF-S} area has been imaged with the Very Large Array (VLA) at 1.4 GHz in two dedicated surveys \citep{kellermann08,miller08}.  Using deep multi-wavelength ancillary data (from the optical to the mid-IR), \citet{mainieri08} find reliable counterparts for 254 ($\sim 95\%$) of the 266 radio sources in these catalogs.  Here, we have acquired spectra and quality redshifts for 48 VLA sources covering a broad range in redshift $0.12 <  z < 3.68$.  A number of these redshifts were presented in \citet{mainieri08}.  We give the full list of identifications (Table~\ref{radio_ids}) and provide optical spectra for the community.  We note that the objects listed in this table are not associated with an X-ray source in either the \citet{le05} or \citet{luo08} catalogs.  We note that there are 18 newly-acquired spectroscopic identifications of X-ray sources that have radio counterparts and are listed in either Table~\ref{opt_catalog} or~\ref{table:2ms_ids}.

\clearpage

\begin{deluxetable}{lllllll}
\tabletypesize{\scriptsize}
\tablecaption{Redshifts for X-ray sources solely detected in the 1-2 Msec catalogs\label{table:2ms_ids}}
\tablewidth{0pt}
\tablehead{
\colhead{XID\tablenotemark{a}}&\colhead{RA$_{\rm opt}$ (J2000)}&\colhead{DEC$_{\rm opt}$ (J2000)} & \colhead{Redshift}&\colhead{Quality}&\colhead{Class}&\colhead{Program}}
\startdata
18 &52.958828& -27.828563 &0.845 &2 &NELG &15\\
31& 52.975166 &-27.834931 &0.744 &2 &NELG&12\\
51& 53.001734 &-27.874613 &0.631 &2 &ALG &10\\
53 &53.003757 &-27.799100 &0.975 &2 &NELG &10\\
92& 53.033335 &-27.782571 &2.619 &2 &NELG &12\\
116& 53.047677 &-27.834926 &2.326 &1& ALG &12\\
130 &53.057590 &-27.757128 &0.643 &1 &NELG &5\\
135& 53.058665 &-27.708438 &2.024 &2 &BLAGN &12\\
136& 53.059319 &-27.779891 &0.840 &2 &NELG &15\\
143& 53.062076 &-27.645422 &0.328 &1 &ALG&5\\
\enddata
\tablenotetext{a}{XID from \citet{luo08}.}
\tablecomments{An abbreviated version of the table is shown.  The full table is provided online with 49 entries.}
\end{deluxetable}

\begin{deluxetable}{llllllll}
\tabletypesize{\scriptsize}
\tablecaption{Spectroscopic redshifts of VLA radio sources\label{radio_ids}}
\tablewidth{0pt}
\tablehead{
\colhead{Radio ID}&\colhead{RA$_{\rm opt}$ (J2000)}&\colhead{DEC$_{\rm opt}$ (J2000)} & \colhead{R$_{AB}$}&\colhead{Redshift}&\colhead{Quality}&\colhead{Class}&\colhead{Program}}
\startdata
5 & 52.81024 & -27.92961 & 23.4 & 1.591 & 2 & BLAGN & KECK\\
12 & 52.83395 & -27.65033 & 20.3 & 0.526 & 2 & NELG & KECK\\
20 & 52.86917 & -27.82635 & 22.8 & 0.846 & 2 & ALG & VLT\\
27 & 52.89256 & -27.64125 & 22.9 & 1.040 & 2 & ALG & KECK\\
34 & 52.93052 & -27.85055 & 19.1 & 0.310 & 2 & NELG & KECK\\
51 & 52.96157 & -27.78431 & 23.6 & 2.259 & 2 & BLAGN & KECK\\
59 & 52.97519 & -27.83488 & 21.9 & 0.744 & 2 & NELG & VLT\\
69 & 52.99545 & -27.71780 & 23.2 & 1.043 & 2 & NELG & KECK\\
75 & 53.01342 & -27.88690 & 24.9 & 1.396 & 9 & NELG & KECK\\
82 & 53.02682 & -27.79132 & 22.7 & 1.021 & 2 & NELG & VLT\\
\enddata
\tablecomments{An abbreviated version of the table is shown.  The full table is provided online with 48 entries.}
\end{deluxetable}
\end{document}